\documentclass[
    ,final            
  ,numberedheadings 
  ]
  {aipproc}

\layoutstyle{6x9}
\usepackage{epsfig,epsf,rotating}
\newcommand{\beq}{\begin{equation}}
\newcommand{\eeq}{\end{equation}}
\newcommand{\beqn}{\begin{eqnarray}}
\newcommand{\eeqn}{\end{eqnarray}}
 
\newcommand\la{\langle}
\newcommand\ra{\rangle}
\newcommand\eps\varepsilon

\def\GeV{\,\mbox{GeV}}

\def\imag{{\rm i}}
\def\euler{{\rm e}}
\def\lsim{\mathrel{\rlap{\lower4pt\hbox{\hskip1pt$\sim$}}
    \raise1pt\hbox{$<$}}}         
\def\gsim{\mathrel{\rlap{\lower4pt\hbox{\hskip1pt$\sim$}}
    \raise1pt\hbox{$>$}}}         
\def\be{\begin{equation}}
\def\ee{\end{equation}}
\def\eq#1{{Eq.~(\ref{#1})}}
\newcommand{\as}{\alpha_s}

\begin{document}

\title{High energy nuclear interactions and QCD: \\ an introduction}

\author{D.E.\ Kharzeev}{
  address={Physics Department, Brookhaven National Laboratory,
Upton, New York 11973, USA}
}

\author{J.\ Raufeisen}{
  address={Los Alamos National Laboratory, MS H846,
Los Alamos, NM 87545, USA}
}

\begin{abstract} 
The goal of these lectures, oriented towards the students just entering the field, 
is to provide an elementary introduction to QCD and 
the physics of nuclear interactions at high energies. 
We first introduce the general structure of QCD and discuss its main 
properties. Then we proceed to Glauber multiple scattering 
theory which lays the foundation for the theoretical treatment of nuclear 
interactions at high energies. We introduce the concept of Gribov's inelastic shadowing, crucial 
for the understanding of quantum formation effects. 
We outline the problems facing Glauber approach at high energies, and discuss 
how asymptotic freedom of QCD helps to resolve them, introducing the 
concepts of parton saturation and color glass condensate.

\end{abstract}

\maketitle




\section{Quantum Chromo-Dynamics -- the theory of strong interactions}

\subsection{What is QCD?}
Strong interaction is, indeed the strongest force of nature. It is
responsible for over 80\% of the baryon masses, and thus for most 
of the mass of everything on Earth. Strong interactions bind nucleons 
in nuclei which, being then dressed with electrons and bound into
molecules by the much weaker electro-magnetic force, give rise
to the variety of the physical world.

Quantum Chromodynamics (QCD) is {\em the} theory of strong interactions. 
The fundamental degrees of freedom of QCD, quarks and gluons, are already 
well established even though they cannot be observed as free particles,
but only in color neutral bound states (confinement). Today,
QCD has firmly occupied its place as part of the Standard Model. However,
understanding the physical world does not only mean understanding its
fundamental constituents; it means mostly understanding how these constituents
interact and bring into existence the entire variety of physical objects
composing the universe. In these lectures, we try to explain
why high energy nuclear physics offers us unique tools to study
QCD. 
  
\subsubsection{The QCD Lagrangian}

So what is QCD? QCD emerges when the na\"{i}ve quark model is combined 
with local SU(3) gauge invariance. Quark model classifies 
the large number of hadrons in terms of a few, more fundamental
constituents. Baryons consist of three quarks, while mesons are made of 
a quark and an antiquark.
For example, the proton is made of
two up-quarks and one down quark, $|p\ra=|uud\ra$, and the $\pi^+$-meson
contains one up and one anti-down quark, $|\pi^+\ra=|u\bar d\ra$. 
However, the quark model in this na\"{i}ve form is not complete, because the
Pauli exclusion principle would not allow for a particle like the $\Delta$ isobar  
$|\Delta^{++}\ra=|uuu\ra$ with spin 3/2. 
The only way to construct a completely
antisymmetric wavefunction for the $\Delta^{++}$ is to postulate
an additional quantum number, which may be called ``color''. 
Quarks can then exist in three different color states; one may choose calling them 
red, green and blue. 
Correspondingly, we can define a quark-state ``vector'' with three components,
\beq
q(x)=\left(
\begin{array}{c}
q^{\rm red}(x)\\
q^{\rm green}(x)\\
q^{\rm blue}(x)
\end{array}
\right).
\eeq
The transition from quark model to QCD is made when one decides to treat color 
similarly to the electric charge in electrodynamics. 
As is well known, the entire structure of electrodynamics emerges from the 
requirement of local gauge invariance, i.e. invariance with respect to the 
phase rotation of electron field, $exp(i \alpha(x))$, where the phase $\alpha$ 
depends on the space--time coordinate.   
One can demand similar invariance for the quark fields, keeping in mind that 
while there is only one electric charge in QED, there are three color charges
in QCD. 

To implement this program, let us require the free quark Lagrangian,
\beq\label{eq:free}
{\cal L}_{{\rm free}}=\sum_{q=u,d,s\dots}\sum_{colors}
\bar q(x)\left(\imag\gamma_\mu\frac{\partial}{\partial x_\mu}-m_q\right)q(x) 
\quad,
\eeq
to be invariant under rotations of the quark fields in color space,
\beq\label{eq:trans}
U:\qquad q^j(x)\quad\to\quad U_{jk}(x)q^k(x),
\eeq
with $j,k\in\{1\dots 3\}$ (we always sum over repeated indices).
Since the theory we build in this way is invariant with respect to these 
``gauge'' transformations, all physically meaningful quantities must be
gauge invariant.

In 
electrodynamics, there is only one electric charge, and 
gauge transformation involves a single phase factor, $U = exp(i \alpha(x))$. 
In QCD, we have three different colors, and $U$ becomes a (complex valued) 
unitary $3\times 3$ matrix, {\em i.e.} $U^\dagger U=UU^\dagger =1$,
with determinant ${\rm Det}\,~U=1$. 
These matrices form the fundamental representation
of the group $SU(3)$ where $3$ is the number of colors, $N_c=3$.
The matrix $U$ has $N_c^2-1=8$ independent elements and can therefore be 
parameterized in terms of the 8 generators ${T}_{kj}^a$, $a\in\{1\dots 8\}$
of the fundamental representation of $SU(3)$,
\beq
U(x) = \exp\left(-\imag\phi_a(x) {T}^a\right)
\eeq
By considering a transformation $U$ that is infinitesimally close to
the $\bf{1}$ element of the group,
it is easy to see that the matrices ${T}^a$ must be Hermitian 
(${T}^{a}={T}^{a\dagger}$) and traceless (${\rm tr}\;{T}^a=0$). 
The ${T}^a$'s do not commute; instead one defines the $SU(3)$ structure
constants $f_{abc}$ by the commutator
\beq
\left[{T}^a,{T}^b\right]=\imag f_{abc}{T}^c.
\eeq
These commutator terms have no analog in QED which is based on the abelian
gauge group $U(1)$. QCD is based on a non-abelian gauge group $SU(3)$ and is 
thus called a non-abelian gauge theory. 

The generators 
${T}^a$ are normalized to
\beq
{\rm tr}\,{T}^a{T}^b=\frac{1}{2}\,\delta_{ab},
\eeq
where $\delta_{ab}$ is the Kronecker symbol. Useful information about
the algebra of color matrices, and their explicit representations, can be found in 
many textbooks (see, e.g., \cite{field}).

Since $U$  
is $x$-dependent, the free quark Lagrangian (\ref{eq:free}) is not
invariant under the transformation (\ref{eq:trans}). In order to preserve 
gauge invariance, one has to introduce, following the familiar case of 
electrodynamics, the gauge (or ``gluon'') field 
$A_{kj}^\mu(x)$ and replace the derivative in (\ref{eq:free}) with the 
so-called {\em covariant derivative},
\beq\label{eq:cov}
\partial^\mu q^j(x)\quad\to\quad
D_{kj}^\mu q^j(x)\equiv\left\{\delta_{kj}\partial^\mu
-\imag A_{kj}^\mu(x)\right\}q^j(x).
\eeq
Note that the gauge field $A_{kj}^\mu(x)=A^{\mu}_a{T}_{kj}^a(x)$ 
as well as the covariant
derivative are $3\times 3$ matrices in color space. Note also that
Eq.~(\ref{eq:cov}) differs from the definition often given  
in textbooks, because we
have absorbed the strong coupling constant in the field $A^\mu$.
With the replacement given by Eq.~(\ref{eq:cov}), 
all changes to the Lagrangian under gauge 
transformations cancel, provided $A^\mu$ transforms as
\beq
U:\qquad
A^\mu(x)\to U(x)A^\mu(x)U^\dagger(x)
+\imag U(x)\partial^\mu U^\dagger(x).
\eeq
(From now on, we will often not write the color indices explicitly.) 

\begin{figure}[b]
  \scalebox{1.0}{\includegraphics*{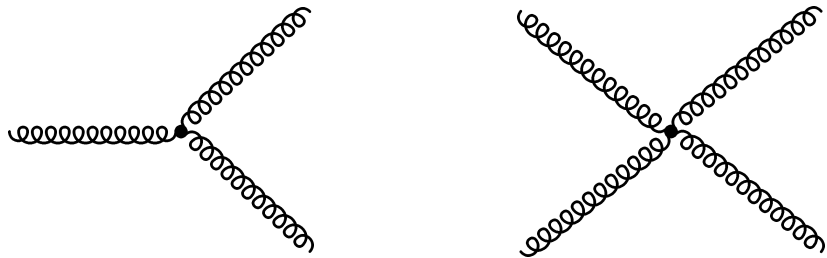}}\hfill
  \raise1cm\hbox{\parbox[b]{2.4in}{
   \caption{\label{fig:gvertex}\em
	Due to the non-abelian nature of QCD, gluons carry color charge
	and can therefore interact with each others via these vertices. 
  	}
  }
}
\end{figure}

The QCD Lagrangian then reads
\beq\label{eq:lag}
{\cal L}_{{\rm QCD}}=\sum_{q}
\bar q(x)\left(\imag\gamma_\mu D^\mu-m_q\right)q(x) 
-\frac{1}{4g^2}{\rm tr}\;G^{\mu\nu}(x)G_{\mu\nu}(x)\quad,
\eeq
where the first term describes the dynamics of quarks and their couplings to
gluons, while the second term describes the dynamics of the gluon field. 
The strong coupling constant $g$ is the QCD analog of the elementary electric
charge $e$ in QED.
The gluon field strength tensor is given by
\beq\label{eq:g}
G^{\mu\nu}(x)\equiv\imag\left[D^\mu,D^\nu\right]
=\partial^\mu A^\nu(x)-\partial^\nu A^\mu(x)
-\imag\left[A^{\mu}(x),A^{\nu}(x)\right].
\eeq
This can also be written in terms of the color components $A_a^{\mu}$
of the gauge field,
\beq
G_a^{\mu\nu}(x)
=\partial^\mu A_a^\nu(x)-\partial^\nu A_a^\mu(x)
+f_{abc}A_b^\mu(x) A_c^\nu(x).
\eeq
For a more complete presentation, see \cite{Guido} and modern textbooks like
\cite{field,Ellis,Muta}.

The crucial, as will become clear soon, difference between electrodynamics 
and QCD is the presence of the commutator on the 
{\em r.h.s.} of Eq.~(\ref{eq:g}).
This commutator gives rise to the gluon-gluon interactions shown in 
Fig.~\ref{fig:gvertex} that make the QCD field equations 
non-linear: the color fields do not simply add like in 
electrodynamics.
These non-linearities give rise to rich and non-trivial dynamics of strong 
interactions.

\subsubsection{Asymptotic Freedom}

\begin{figure}
\caption{\label{fig:qedscreen}\em
     In QED, virtual electron-positron pairs from the vacuum screen
the bare charge of the electron. The larger the distance, the more pairs
are present to screen the bare charge and the electromagnetic coupling
decreases. Conversely, the coupling is larger when probed at short 
distances.}
\includegraphics[height=.3\textheight]{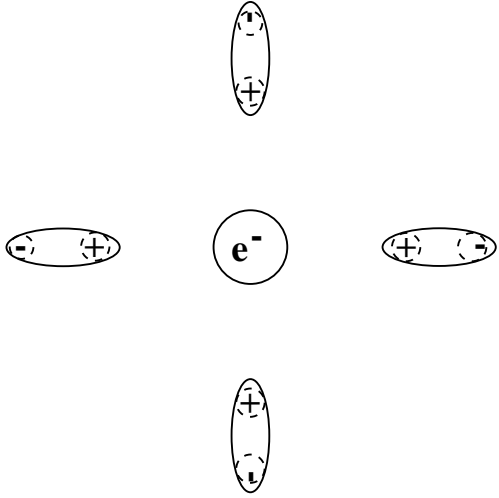}
\end{figure}

Let us now turn to the discussion of the dynamical properties of QCD. To 
understand the dynamics of a field theory, one necessarily has 
to understand how 
the coupling constant behaves as a function of distance. 
This behavior, in turn, is determined by the response of the 
vacuum to the presence of external charge. The vacuum
is the ground state of the theory; however, quantum mechanics tells us that 
the ``vacuum'' 
is far from being empty -- 
the uncertainty principle 
allows particle-antiparticle pairs to be present in the vacuum for a period time 
inversely proportional to their energy. In QED,
the electron-positron pairs have the effect of screening the electric charge,
see Fig.~\ref{fig:qedscreen}. Thus, the electromagnetic coupling constant 
increases toward shorter distances. The dependence of the charge on distance 
is given by 
\beq\label{eq:eofr}
e^2(r)=\frac{e^2(r_0)}{1+\frac{2e^2(r_0)}{3\pi}\ln\frac{r}{r_0}},
\eeq
which can be obtained by resumming (logarithmically divergent, and regularized at the 
distance $r_0$) electron--positron loops dressing the virtual photon propagator. 

The formula (\ref{eq:eofr}) has two surprising properties: first,  
at large distances $r$ away from the charge which is 
localized at $r_0$, $r\gg r_0$, where one can neglect 
unity in the denominator, 
the ``dressed'' charge $e(r)$ becomes independent of the value of the ``bare'' charge $e(r_0)$ -- 
it does not matter what the value of the charge at short distances is.  
Second, in the local 
limit $r_0\to 0$, if we require the bare charge $e(r_0)$ be finite, 
the effective charge vanishes at any finite distance away 
from the bare charge! This is the celebrated Landau's zero charge problem \cite{zero}: 
the screening of the charge in QED does not allow to reconcile the presence 
of interactions with the local limit of the theory. 
This is a fundamental problem of QED, which shows that i) either it is not a truly 
fundamental theory, or ii) Eq. (\ref{eq:eofr}), based on perturbation theory, in the strong coupling 
regime gets replaced 
by some other expression with a more acceptable behavior. The latter possibility is 
quite likely since at short distances the electric charge becomes very large and 
its interactions with electron--positron vacuum cannot be treated perturbatively. 
A solution of the zero charge problem, based on considering the rearrangement of 
the vacuum in the presence of ``super--critical'', at short distances, charge was 
suggested by Gribov \cite{gribov}.
 
Fortunately, because of the smallness
of the physical coupling $\alpha_{em}(r)=e^2(r)/(4\pi)=1/137$, this 
fundamental problem of the theory manifests itself only at very short distances 
$\sim\exp(-3/[8\alpha_{em}])$. Such short distances will probably 
always remain beyond the reach of experiment, and one can safely apply
QED as a truly effective theory.

\vskip0.3cm

In QCD, as we are now going to discuss, the situation is qualitatively 
different, and corresponds to {\it anti-}screening -- 
the charge is small at short distances and grows at larger distances. 
This property of the theory, discovered by Gross, Wilczek, and Politzer \cite{alphas}, 
is called asymptotic freedom. 

While the derivation of the running coupling is conventionally performed 
by using field theoretical perturbation theory, it is instructive to see how these  
results can be illustrated by using the methods of condensed matter physics. 
Indeed, let us consider the vacuum as a continuous medium
with a dielectric constant $\epsilon$. 
The dielectric constant is linked to the magnetic permeability $\mu$ and the
speed of light $c$ by the relation
\beq
\epsilon\,\mu=\frac{1}{c^2}=1.
\eeq
Thus, a screening medium ($\epsilon>1$) will be diamagnetic ($\mu<1$),
and conversely a paramagnetic medium ($\mu>1$) will exhibit antiscreening
which leads to asymptotic freedom. In order to calculate the running
coupling constant, one has to calculate the magnetic permeability 
of the vacuum. We follow \cite{nielsen} in our discussion, where
this has been done  in a framework very similar to
Landau's theory of the diamagnetic properties of a free electron gas.
In QED one has
\beq
\epsilon_{QED}=1+\frac{2e^2(r_0)}{3\pi}\ln\frac{r}{r_0}>1
\eeq
So why is the QCD vacuum paramagnetic while the QED vacuum is diamagnetic?
The energy density of a medium in the presence of an external magnetic field 
$\vec B$ is given by
\beq\label{eq:vac}
u=-\frac{1}{2}4\pi\chi\vec B^2
\eeq
where the magnetic susceptibility $\chi$ is defined by the relation
\beq\label{eq:chi}
\mu=1+4\pi\chi.
\eeq
When electrons move in an external magnetic field, two competing effects determine
the sign of magnetic susceptibility:
\begin{itemize}
\item{The electrons in magnetic field move along quantized orbits, so-called Landau levels.
The current originating from this movement produces a magnetic field 
with opposite direction to the external field. This is the diamagnetic
response, $\chi<0$.}
\item{The electron spins align along the direction
of the external $\vec B$-field, leading to a paramagnetic response ($\chi>0$).}
\end{itemize}
In QED, the diamagnetic effect is stronger, so the vacuum is screening the bare
charges. In QCD, however, gluons carry color charge. Since they have a larger
spin (spin 1) than quarks (or electrons), the paramagnetic effect dominates and
the vacuum is anti-screening.  

Let us explain this in more detail. Basing on the considerations given above, 
the energy density of the QCD vacuum 
in the presence of an external color-magnetic field
can be calculated by using the standard formulas of quantum mechanics, 
see {\em e.g.} \cite{ll3}, by summing over 
Landau levels  
and taking account of the fact that gluons and quarks give contributions 
of different sign. Note that a summation over all Landau levels would lead to an infinite result
for the energy density. In order to avoid this
divergence, one has to introduce a cutoff $\Lambda$ with dimension of mass.
Only field modes with wavelength $\lambda\gsim 1/\Lambda$ are taken into account.
The upper limit for $\lambda$ is given by the radius of the largest Landau orbit, 
$r_0\sim 1/\sqrt{gB}$, which is the only dimensionful scale in
the problem; the summation thus is made over the wave lengths satisfying
\beq\label{eq:limits}
\frac{1}{\sqrt{|gB|}}\gsim\lambda\gsim\frac{1}{\Lambda},
\eeq
The result is \cite{nielsen} 
\beq\label{eq:edens}
u_{vac}^{QCD}=-\frac{1}{2}B^2\frac{11N_c-2N_f}{48\pi^2}g^2
\ln\frac{\Lambda^2}{\left|gB\right|},
\eeq
where $N_f$ is the number of quark flavors, and $N_c=3$ is the number of flavors. 
Comparing this with Eqs.~(\ref{eq:vac}) and (\ref{eq:chi}), one can read
off the magnetic permeability of the QCD vacuum,
\beq
\mu_{vac}^{QCD}(B)=1+\frac{11N_c-2N_f}{48\pi^2}g^2
\ln\frac{\Lambda^2}{\left|gB\right|}>1.
\eeq
The first term in the denominator ($11N_c$) is the gluon contribution
to the magnetic permeability. This term dominates over the quark contribution
($2N_f$) as long as the number of flavors $N_f$ is less than 17 and is responsible
for asymptotic freedom.

\begin{figure}[b]
  \scalebox{0.45}{\includegraphics*{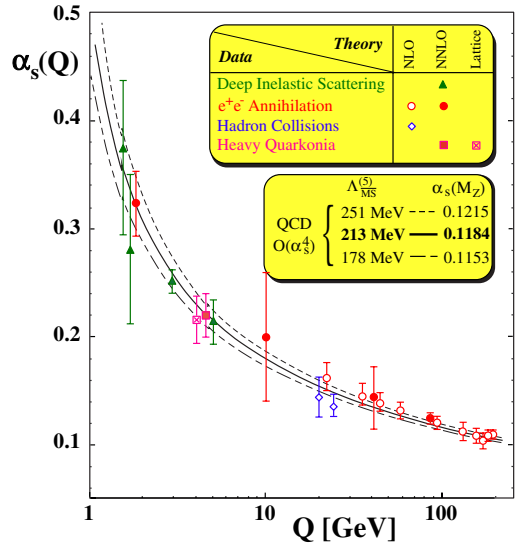}}\hfill
  \raise1.1cm\hbox{\parbox[htb]{3in}{
   \caption{\label{fig:bethke}\em
	The running coupling constant $\alpha_s(Q^2)$ as a function
of momentum transfer $Q^2$ determined from a variety of processes.
The figure is from \cite{Bethke}, courtesy of S. Bethke.
  	}
  }
}
\end{figure}


The dielectric constant as a function of distance $r$ is then given
by
\beq\label{eq:em}
\epsilon_{vac}^{QCD}(r)=\left. 
\frac{1}{\mu_{vac}^{QCD}(B)}\right|_{\sqrt{|gB|}\to 1/r}.
\eeq
The replacement $\sqrt{|gB|}\to 1/r$ follows from the fact that 
$\epsilon$ and $\mu$ in Eq.~(\ref{eq:em}) should be calculated from the same
field modes: the dielectric constant $\epsilon(r)$ could be calculated by 
computing the vacuum energy in the presence of two static colored test particles 
located at a distance $r$ from each other. In this case, the maximum wavelength
of field modes that can contribute is of order $r$ so that
\beq\label{eq:limits2}
r\gsim\lambda\gsim\frac{1}{\Lambda}.
\eeq
Combining Eqs.~(\ref{eq:limits}) and 
(\ref{eq:limits2}), we identify $r= {1}/{\sqrt{|gB|}}$ and find
\beq
\epsilon_{vac}^{QCD}(r)=\frac{1}{1+\frac{11N_c-2N_f}{24\pi^2}g^2
\ln(r\Lambda)}<1.
\eeq
With $\alpha_s(r_1)/\alpha_s(r_2)
=\epsilon_{vac}^{QCD}(r_2)/\epsilon_{vac}^{QCD}(r_1)$ 
one finds to lowest order in $\alpha_s$ 
\beq
\alpha_s(r_1)=\frac{\alpha_s(r_2)}{1+\frac{11N_c-2N_f}{6\pi}\alpha_s(r_2)
\ln\!\left(\frac{r_2}{r_1}\right)}. \label{asfr1}
\eeq
Apparently, if $r_1<r_2$ then $\alpha_s(r_1)<\alpha_s(r_2)$. The running of the 
coupling constant is shown in Fig.~\ref{fig:bethke}, $Q\sim 1/r$.
The intuitive derivation given above illustrates the original field--theoretical 
result of \cite{alphas}.

At high momentum transfer, corresponding to short distances,
the coupling constant
thus becomes small and one can apply perturbation theory, see Fig.~\ref{fig:bethke}.
There is a variety of processes that involve high momentum scales, {\em e.g.}
deep inelastic scattering, Drell-Yan dilepton production, $e^+e^-$-annihilation
into hadrons, production of heavy quarks/quarkonia,
high $p_T$ hadron production $\dots$. QCD correctly predicts the $Q^2$ dependence of these,  
so-called ``hard'' processes, which is a great success of the theory.

\subsection{Challenges in QCD}

\subsubsection{Confinement}

While asymptotic freedom implies that the theory becomes simple and treatable at 
short distances, it also tells us that at large distances the coupling becomes 
very strong. In this regime we have no reason to believe in perturbation theory. 
In QED, as we have discussed above, the strong coupling regime starts at 
extremely short distances beyond the reach of current experiments -- and this 
makes the ``zero--charge'' problem somewhat academic. In QCD, the entire physical 
World around us is defined by the properties of the theory in the strong coupling 
regime -- and we have to construct accelerators to study it in the much more simple, 
``QED--like'', weak coupling limit.

We do not have to look far to find the striking differences between the properties 
of QCD at short and large distances: 
the elementary building blocks of QCD -- the ``fundamental'' fields 
appearing in the Lagrangean (\ref{eq:lag}), quarks and gluons, do not exist in the physical spectrum 
as asymptotic states. For some, still unknown to us, reason, all physical states 
with finite energy appear to be color--singlet combinations of quarks and gluons, which 
are thus always ``confined'' at rather short distances on the order of 1 fm. 
This prevents us, at least in principle, from using well--developed 
formal S-matrix approaches based on analyticity and unitarity to describe quark and gluon  
interactions. 

The property of confinement can be explored by looking at the 
propagation of heavy quark--antiquark pair at a distance $R$ propagating in time 
a distance $T$. An object which describes the behavior of this system is the Wilson loop \cite{Wilson}
\beq 
W(R,T) = Tr \left[P\ exp\left[i\int_C A_{\mu}^a T^a dx^{\mu}\right]\right],
\eeq
where $A_{\mu}^a$ is the gluon field, $T^a$ is the generator of $SU(3)$, 
and the contour $C$ is chosen as a rectangle 
with side $R$ in one of the space dimensions and $T$ in the time direction. 
It can be shown that at large $T$ the asymptotics of the Wilson loop is
\beq
\lim_{T \to \infty}\ W(R,T) = exp\left[- T V(R)\right], 
\eeq
where $V(R)$ is the static potential acting between the heavy quarks. At large distances, 
this potential grows as 
\beq
V(R) = \sigma R, \label{string} 
\eeq
where $\sigma \sim 1 \ {\rm GeV}/{\rm fm}$ is 
the string tension.  We thus conclude that at large $T$ and $R$ the Wilson loop should 
behave as
\beq
W(R,T) \simeq exp\left[- \sigma T R\right], \label{area}
\eeq
The formula (\ref{area}) is the celebrated ``area law'', which 
signals confinement. 

It should be noted, however, that the introduction of dynamical quarks 
leads to the string break--up at large distances, and the potential $V(R)$ saturates 
at a constant. The presence of light dynamical quarks is most important in Gribov's 
confinement scenario \cite{gribov}, in which the color charges at large distances behave similarly 
to the ``supercritical'' charge in electrodynamics, polarizing the vacuum and 
producing copious quark--antiquark pairs which screen them. In this scenario, 
in the physical world with light quarks there is never a confining force acting 
on color charges at large distances, just quark--antiquark pair production 
(``soft confinement''). 
This may explain why the spectra of jets, for example, computed in perturbative QCD, 
appear to be consistent with experiment; this fact would be difficult to reconcile with the 
existence of strong confining forces. 
There exists a  special situation, however, when the law (\ref{area}) should be 
appropriate even in the presence of light quarks -- the heavy quarkonium. 
The sizes of heavy quarkonia are quite small, and their masses are 
below the threshold to produce a pair of heavy mesons. This is why heavy quarkonia 
are especially useful probes of confinement. 

At high temperatures, the long--range interactions responsible for confinement become  
screened away -- instead of the growing potential (\ref{string}), we expect 
\beq
V(R) \sim - {g^2(T) \over R}\ exp(-m_D R),
\eeq
where $m_D \sim g T$ is the Debye mass. Mathematically, this transition to the 
deconfined phase can again be studied by looking at the properties of the Wilson loop. 
At finite temperature, the theory is defined on a cylinder: 
Euclidean time $\tau$ 
varies within $0 \leq \tau \leq \beta = 1/T$, and the gluon fields satisfy the periodic boundary 
conditions:
\beq 
A_{\mu}^a(\vec{x}, 0) = A_{\mu}^a(\vec{x}, \beta).
\eeq   
Let us now consider the Wilson loop wrapped around this cylinder (the Polyakov loop), 
and choose a gauge where $A_0^a$ is time--independent:
\beq
P({\vec x}) = Tr\ exp\left[ i g \beta A_0^a(\vec{x}) t^a \right];
\eeq
the correlation function of these objects can be defined as
\beq
C_T({\vec x}) = < P(\vec{x}) P^*(\vec{x}) >_T.
\eeq
Again, it can be shown that this correlation function 
is related to the free energy, and thus static potential $V(R)$, 
of the heavy quark--antiquark pair.
Assuming, as before, that the heavy quarks are separated by the spatial distance $R=|\vec{x}|$, 
one finds
\beq
C_T(R) \sim exp \left[- \beta V(R)\right].
\eeq
Again, if we define the limit value $L(T)$ of the correlation function,  
\beq
\lim_{R \to \infty}\ C_T(R) \equiv L(T)
\eeq
it would have to vanish in the confined phase in the absence of dynamical quarks, since 
$V(R)$ tends to infinity in this case: $L(T) = 0$. In the deconfined phase, on the other 
hand, because of the screening $V(R)$ should tend to a constant, and this implies a finite 
value $L(T) \neq 0$. The correlation function of Polyakov loops therefore can be used 
as an order parameter of the deconfinement. The behavior of $L(T)$ as a function of 
temperature has been measured on the lattice; one indeed observes a transition from the 
confined phase with $L(T) = 0$ to the deconfined phase with $L(T) \neq 0$ at some 
critical temperature $T_c$.  
In the presence of light quarks, as we have 
already discussed above, the potential would tend to a constant even in the confined 
phase, and $L(T)$ ceases to be a rigorous order parameter.

\subsubsection{Chiral symmetry breaking}

The decades of experience with ``soft pion'' techniques and current algebra 
convinced physicists that the properties of the 
world with massless pions are quite close to the properties of our physical World. 
The existence of massless particles is always a manifestation of a symmetry of the theory -- 
photons, for example, appear as a consequence of local gauge invariance of the 
electrodynamics.  
However, unlike photons, pions have zero spin and cannot be gauge bosons of any symmetry. The 
other possibility is provided by the Goldstone theorem, which states that the appearance 
of massless modes in the spectrum can also reflect a spontaneously broken symmetry, 
i.e. the symmetry of the theory which is broken in the ground state. Because of the 
great importance of this theorem, let us briefly sketch its proof. 

Suppose that the 
Hamiltonian $H$ of the theory is invariant under some symmetry generated by operators $Q_i$, 
so that 
\beq
\left[H, Q_i\right] = 0.
\eeq
Spontaneous symmetry breaking in the ground state of theory implies that for some of the 
generators $Q_i$
\beq 
Q_i |0> \neq 0.
\eeq   
Since $Q_i$ commute with the Hamiltonian, this means that this new state $Q_i |0>$ 
has the same energy as the ground state. The vacuum is therefore degenerate, and 
in a relativistically invariant theory this implies the existence of massless 
particles -- Goldstone bosons.
A useful example of that is provided by the  phonons in a crystal, where the continuous 
translational symmetry of the QED Lagrangean is spontaneously broken by the existence of 
the fixed period of the crystal lattice. 

Even though all six quark flavors enter the Lagrangean, it is intuitively 
clear that at small scales $Q << M_c, M_b, M_t$, heavy quarks should not have 
any influence on the dynamics. In a rigorous way this statement is formulated in terms 
of decoupling theorems, which we will discuss in detail later. At the moment let us 
just assume that we are interested in the low--energy behavior, and that only light 
quarks are relevant for that purpose. Then it makes sense to consider the approximate 
symmetry, which becomes exact when the quarks are massless. In fact, in this limit, 
the Lagrangean does not contain any terms which connect the right-- and left--handed 
components of the quark fields:
\beq 
q_R ={1 \over 2} (1 + \gamma_5) q; \hskip1cm q_L ={1 \over 2} (1 - \gamma_5) q.
\eeq
The Lagrangean of QCD (\ref{eq:lag}) is therefore invariant under the independent transformations of 
right-- and left--handed fields (``chiral rotations''). In the limit of massless quarks, 
QCD thus possesses 
an additional symmetry $U_L(N_f) \times U_R(N_f)$ 
with respect to the independent transformation of left-- and right--handed 
quark fields $q_{L,R} = {1 \over 2}(1 \pm \gamma_5) q$: 
\beq
q_L \to V_L q_L; \ \ q_R \to V_R q_R; \ \ V_L, V_R \in U(N_f); \label{chiral}
\eeq
this means that left-- and right--handed quarks are not correlated.  

Even a brief look into the Particle Data tables, or simply in the 
mirror, can convince anyone 
that there is no symmetry between left and right in the physical World. 
One thus has to assume that the symmetry (\ref{chiral}) is spontaneously 
broken in the vacuum. 
 
The presence of the ``quark condensate'' $< \bar{q}q >$ in QCD vacuum signals spontaneous 
breakdown of this symmetry, since 
\beq
< \bar{q}q > = < \bar{q_L}q_R > + < \bar{q_R}q_L >,
\eeq
which means that left-- and right--handed quarks and antiquarks can transform 
into each other. Quark condensate therefore can be used as an order parameter 
of chiral symmetry. Lattice calculations show that around the deconfinement phase 
transition, quark condensate dramatically decreases, signaling the onset of the 
chiral symmetry restoration. 

This spontaneous breaking of $U_L(3) \times U_R(3)$ chiral symmetry, 
by virtue of the Goldstone theorem presented above, should give rise 
to $3^2 = 9$ Goldstone particles. 
The flavor composition of the existing eight candidates for this role 
(3 pions, 4 kaons, and the $\eta$) suggests that the 
$U_A(1)$ part of $U_L(3) \times U_R(3) = 
SU_L(3) \times SU_R(3) \times U_V(1) \times U_A(1)$ does not exist.  
This constitutes the famous ``$U_A(1)$ problem''. 

\subsubsection{The origin of mass}

There is yet another problem with the chiral limit in QCD. Indeed, as the 
quark masses are put to zero, the Lagrangian (\ref{eq:lag}) does not contain 
a single dimensionful scale -- the only parameters are pure numbers $N_c$ 
and $N_f$. The theory is thus apparently 
invariant with respect to scale transformations, 
and the corresponding scale current is conserved: 
$\partial_{\mu} s_{\mu} = 0$.      
However, the absence of a mass scale would imply that all physical 
states in the theory should be massless!

\subsubsection{Quantum anomalies}

Both apparent problems -- the missing $U_A(1)$ symmetry and the 
origin of hadron masses -- are related to quantum anomalies.
A symmetry of a classical theory can be broken when that theory
is quantized, due to the requirements of regularization and renormalization.
This is called anomalous symmetry breaking. Regularization of the theory 
on the quantum level brings in a dimensionful parameter -- remember 
the cutoff $\Lambda$ of Eq. (\ref{eq:limits}) we had to impose 
on the wavelength of quarks and gluons. 

Once the theory is quantized, we already 
know that the coupling constant is scale dependent and therefore
scale invariance is broken (note that the four-divergence of the scale current 
in field theory is equal
to the trace of the energy momentum tensor $\Theta^\mu_{\;\mu}$). One finds
\beq\label{eq:trace}
\partial^\mu s_\mu=\Theta^\mu_{\;\mu}=\sum_qm_q\bar qq+\frac{\beta(g)}{2g^3}
{\rm tr G^{\mu\nu}G_{\mu\nu}},
\eeq
where $\beta(g)$ is the QCD $\beta$-function, which governs the behavior of the 
running coupling: 
\beq
\mu {d g(\mu) \over d \mu} = \beta (g); \label{rg}
\eeq   
note that as discussed in Section {\it 1.1.1} 
we include coupling $g$ in the definition of the gluon fields. 
As we already discussed, at small coupling $g$, the $\beta$ function is negative, which means that 
the theory is asymptotically free. The leading term in the perturbative 
expansion is (compare with Eq. (\ref{asfr1}))
\beq
\beta(g) = -b {g^3 \over (4\pi)^2} + O(g^5), \hskip1cm b = 11 N_c - 2 N_f, \label{betafu}
\eeq  
where $N_c$ and $N_f$ are the numbers of colors and flavors, respectively. 

\begin{figure}[t]
\centerline
{
  \scalebox{1.0}{\includegraphics{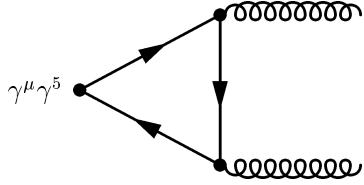}}
 }
{\parbox[b]{13cm}{
\caption{\label{fig:triangle}\em
 The triangle graph that leads to the $U_A(1)$-anomaly. The corresponding
graph with the two gluons interchanged in the final state is not shown.
 }}}
\end{figure}

Hadron masses are related
to the forward matrix element of trace of the QCD energy-momentum tensor,
$2m_h^2=\la h|\Theta^\mu_{\;\mu}|h\ra$. Apparently, light hadron masses must receive 
dominant contributions from the $G^2$-term in Eq.~(\ref{eq:trace}). Note also 
that the flavor sum in Eq.~(\ref{eq:trace}) includes heavy flavors, too. 
This would lead to the unphysical picture that {\em e.g.} the proton mass
is dominated by heavy quark masses. However, the heavy flavor contribution
to the sum (\ref{eq:trace}) is exactly canceled by a corresponding  
heavy flavor contribution to the $\beta$-function.

Similar thing  
happens with the
axial current, $j^5_\mu=\bar q\gamma_\mu\gamma^5q$, generated by the $U_A(1)$ group. 
The corresponding axial charge is not
conserved because of the contribution of the triangle graph in Fig.~\ref{fig:triangle}, and
the four-divergence of the axial current is given by
\cite{abj}
\beq
\partial^\mu j_\mu^5=\sum_q2\imag m_q\bar q\gamma^5q +\frac{N_f}{8\pi^2}\,
{\rm tr}\,G^{\mu\nu}\widetilde G_{\mu\nu}, \label{axan}
\eeq  
where $\widetilde G_{\mu\nu}=\epsilon_{\mu\nu\kappa\lambda}G^{\kappa\lambda}/2$
is the dual field strength tensor.
Since the gluonic part on the {\em rhs} of this equation is a surface term (a full divergence),  
there would be no physical effect, if the QCD vacuum were  
 ``empty''. 

\subsubsection{Classical solutions}

\begin{figure}[t]
\centerline
{
  \scalebox{0.6}{\includegraphics{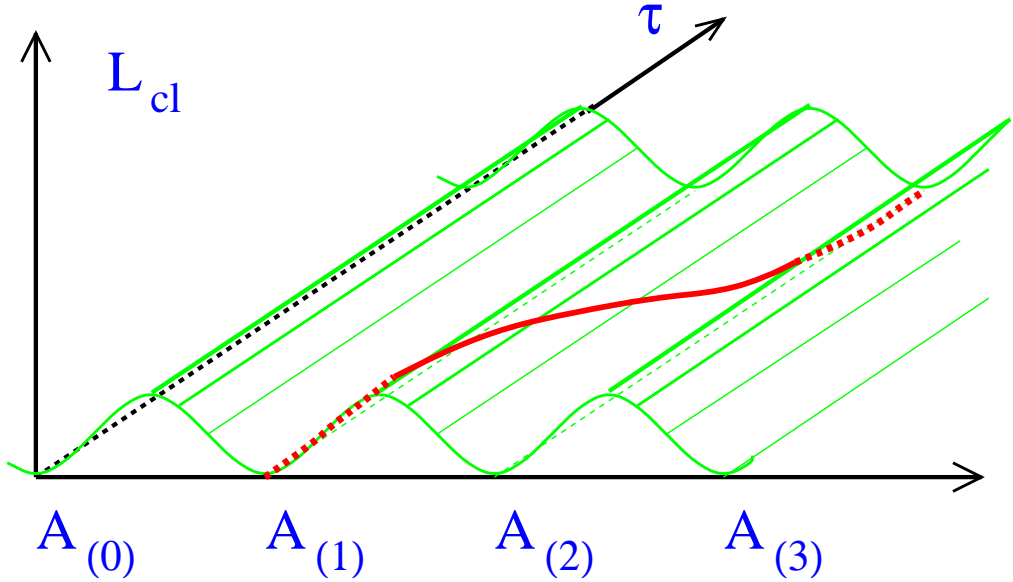}}
 }
{\parbox[b]{13cm}{
\caption{\label{fig:vacuum}\em
Topological structure of QCD vacuum.
  The minima correspond to classical ground states with
topologically different gauge field configurations $A_{(n)}$. 
Also shown is an instanton trajectory interpolating
between the classical vacua $A_{(1)}$ and  $A_{(2)}$. The third axis shows 
the Euclidean time $\tau$. From \cite{hilmar}; courtesy of H. Forkel. 
 }}}
\end{figure}

However, it appears that due to non--trivial topology of the $SU(3)$ 
gauge group, QCD equations of motion allow classical solutions even 
in the absence of external color source, i.e. in the vacuum. 
The well--known example of a classical solution is the {\it instanton}, 
corresponding 
to the mapping of a three--dimensional sphere $S^3$ onto the $SU(2)$ subgroup 
of color $SU(3)$ (for reviews, see \cite{shuryak, hilmar}). 
As a result, the ground state of classical Chromodynamics is not unique. 
There is an enumerable
infinite number of gauge field configurations with different 
topologies (corresponding to different winding number in the $S^3 \to SU(2)$ mapping), 
and the ground state looks like a periodic potential,
see Fig.~\ref{fig:vacuum}. 

In a quantum theory, however, the system will not
stay in one of the minima, like the classical system would. Instead, there 
will be tunneling processes between different minima. These tunneling processes, 
in Minkowski space, correspond to instantons. Since tunneling, in general, lowers 
the ground state energy of the 
system, one expects the QCD vacuum to have a complicated structure.

Instantons, through the anomaly relation (\ref{axan}), lead to the explicit 
violation of the $U_A(1)$ symmetry and thus solve the mystery of the missing 
ninth Goldstone boson - the $\eta'$. Physically, axial symmetry $U_A(1)$ is broken because the 
tunneling processes between topologically different vacua are accompanied by the change 
in quark helicity -- even in the vacuum, left-handed 
quarks periodically turn into right-handed and 
{\it vice versa}. 

\subsubsection{Strong CP problem}

The vacuum structure shown in Fig.~\ref{fig:vacuum} immediately leads to a 
puzzle known as the {\em strong CP problem:} When one calculates 
the expectation
value of an observable in the vacuum, one has to average over all topological
sectors of the vacuum. This is equivalent to adding an additional 
term to the QCD-Lagrangian,
\beq\label{eq:theta}
{\cal L}_{\rm QCD} \to {\cal L}_{\rm QCD}
-\frac{\theta}{16\pi^2}\;{\rm tr}\;G^{\mu\nu}\widetilde G_{\mu\nu},
\eeq
where $\theta\in [0,2\pi]$ 
is a parameter of the theory 
which has to be determined from experiment. Since the $\theta$-term in 
Eq.~(\ref{eq:theta}) is CP violating, a non-zero value of $\theta$ would have 
immediate phenomenological consequences, {\em e.g.} an electric dipole moment
of the neutron.
However, precision measurements of this
dipole moment constrain $\theta$ to  
$\theta<10^{-9}$. The fact that $\theta$ is so unnaturally small 
constitutes the strong CP problem. The most likely solution to this problem 
\cite{axion} implies the existence of a light pseudoscalar meson, the 
{\em axion}. However, despite many efforts, axions remain unobserved in
experiment. 

\subsubsection{Phase structure}

As was repeatedly stated above, the most important problem facing us in 
the study of all aspects of QCD is understanding the structure of the 
vacuum, which, in a manner of saying, does not at all behave as 
an empty space, but as a physical entity with a complicated structure. 
As such, the vacuum can be excited, altered and modified in physical 
processes \cite{leewick}.

Collisions of heavy ions are the best way to create high energy density 
in a ``macroscopic'' (on the scale of a single hadron) volume. 
 It thus could be possible to create and to study a new state of 
matter,
the {\em Quark-Gluon Plasma}(QGP), 
in which quarks and gluons are no longer 
confined in hadrons, but can propagate freely. The search for
QGP is one of the main motivations for the heavy ion research. 

Lattice calculations predict that QCD at high temperatures undergoes phase transitions 
in which confinement property is lost and chiral symmetry is restored. The critical 
temperature for the chiral phase transition is similar (or maybe even equal)
to the critical temperature for deconfinement.
 
Heavy ion collisions at RHIC may also give us the possibility to study the $\theta$ angle 
dependence of the QCD phase diagram. In a heavy ion collision, bubbles 
containing a metastable vacuum with $\theta\neq 0$ may be produced, 
and reveal themselves through their unusual decay pattern \cite{cp}.

\section{Nuclear interactions at high energies}

\subsection{Glauber-Gribov Theory}

It is intuitively clear that heavy ion collisions are governed
by multiple scattering effects. As a short introduction to the
basics of multiple scattering theory,
we introduce here the eikonal approximation to high
energy scattering processes and the Glauber multiple scattering theory 
\cite{Glauber}. We also discuss Gribov's inelastic corrections 
\cite{glaubergribov}
to Glauber's theory.

\subsubsection{The Eikonal Approximation}

The eikonal approximation is the classical approximation to the angular
momentum $l$. In partial wave expansion, {\em i.e.} in an expansion
in angular momentum eigenstates, the scattering amplitude $f(s,t)$ reads
\cite{ll3}
\beq\label{eq:fl}
f(s,t)=\frac{1}{2\imag p}\sum_l(2l+1)
\left[\euler^{2\imag\delta_l}-1\right]
{\rm P}_l(\cos\theta),
\eeq
where $s$ and $t$ are the usual Mandelstam variables (center-of-mass energy squared and invariant 
momentum transfer, respectively), $p$ is the momentum of
the projectile and P$_l$ are the Legendre functions, which depend on the 
cosine of the scattering angle $\theta$. All information about 
the interaction is contained in the scattering phases $\delta_l$.

High energy scattering is of course a process that is far from being
spherically symmetric. Therefore, very large values of $l$ will dominate the
sum Eq.~(\ref{eq:fl}) and we can treat the angular momentum classically.
Since the angular momentum is given by $pb$, one replaces the variable $l$
by the impact parameter $b$, 
\beq
pb=l+\frac{1}{2}.
\eeq
Note that $b$ is now a continuous variable, so angular momentum is no
longer quantized.

At large $l$ and for small scattering angles $\theta$, the Legendre functions
can be expressed to good approximation as
\beq\label{eq:plk}
{\rm P}_l(\cos\theta)\approx\int_0^{2\pi}\frac{d\phi}{2\pi}
\euler^{\imag(2l+1)\sin(\theta/2)\cos(\phi)}=
\int_0^{2\pi}\frac{d\phi}{2\pi}
\euler^{\imag\vec q\cdot\vec b},
\eeq
where $\vec q=\vec p-\vec p^\prime$ is the momentum transfer in the 
scattering process ($t\approx-|\vec q|$)
and $|\vec p|=|\vec p^\prime|$ for elastic scattering. 
At high energy, $\vec q$ lies in the impact parameter plane.
We have used the relation 
\beq
(2l+1)\sin(\theta/2)\cos(\phi)=2p\sin(\theta/2)\frac{l+1/2}{p}\cos(\phi)
=\vec q\cdot\vec b
\eeq
to obtain the second equality in Eq.~(\ref{eq:plk}).

Thus, the scattering amplitude in eikonal approximation reads
\beq
f(s,t)=\frac{\imag p}{2\pi}\int d^2b\euler^{\imag\vec q\cdot\vec b}
\left[1-\euler^{\imag\chi(s,\vec b)}\right],
\eeq
where the phase shift of the projectile is related to the scattering phase
$\delta_l$ by
\beq
\chi(s,\vec b)\equiv 2\delta(s,b).
\eeq
In the case of scattering off a potential $V(\vec r)$, 
this phase shift is simply given by 
\beq
\chi(\vec b)=-\frac{1}{v}\int_{-\infty}^\infty V(\vec r)dz,
\eeq
where $v$ is the velocity of the projectile. The scattering amplitude then
reads
\beq
f(s,t)=\frac{\imag p}{2\pi}\int d^2b\euler^{\imag\vec q\cdot\vec b}
\left[1-\exp\left(-\frac{\imag}{v}\int V(\vec r)dz\right)\right].
\eeq
The total cross section can now be obtained from the forward scattering
amplitude via the optical theorem,
\beq\label{eq:tot}
\sigma_{tot}=\frac{4\pi}{p}{\rm Im}\,f(s,t=0)=
2\int d^2b\left(1-{\rm Re}\,\euler^{\imag\chi(b)}\right).
\eeq

For completeness, we also give the expressions for the elastic and inelastic
cross sections. The elastic cross section is obtained by squaring 
the elastic scattering amplitude and integrating over the solid angle,
\beq
\sigma_{el}=\int d\Omega_{p^\prime}\left|f(\theta,\phi)\right|^2.
\eeq
With the approximation $d\Omega_{p^\prime}\approx d^2p^\prime/p^{\prime 2}$,
which assumes that scattering takes place predominantly in forward direction,
one obtains
\beq\label{eq:el}
\sigma_{el}=
\int d^2b\left|1-\euler^{\imag\chi(b)}\right|^2
\eeq
Finally, the inelastic cross section is
\beq
\label{eq:inel}\sigma_{inel}=\sigma_{tot}-\sigma_{el}=
\int d^2b\left(1-\left|\euler^{\imag\chi(b)}\right|^2\right).
\eeq
For potential scattering, the inelastic cross section, of
course, vanishes because $\chi(b)$ is real. In general, however, $\chi(b)$ 
will have an imaginary part.

The expressions Eqs.~(\ref{eq:tot}), (\ref{eq:el}) and (\ref{eq:inel})
could have been obtained directly from the partial wave decomposition
of the total, elastic and inelastic cross section, as well.
The conditions under which the eikonal approximation is applicable
are investigated in detail in \cite{Glauber}.

\subsubsection{Multiple Scattering Theory}

Based on the eikonal approximation, it is quite straightforward to
develop a theory for scattering off a composite system. 
In this section, we explain the basic features of the multiple scattering
theory developed by Glauber \cite{Glauber}. A much more detailed
presentation of this subject can be found in \cite{Glauber}.

\begin{figure}[t]
\centerline
{
  \scalebox{0.8}{\includegraphics{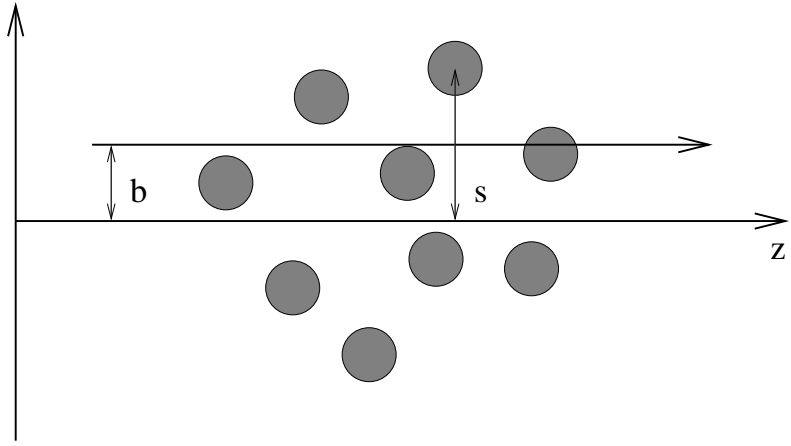}}
 }
{\parbox[b]{13cm}{
\caption{\label{fig:mscat}\em Scattering off a composite system. 
The impact parameter of the projectile is denoted by $\vec b$, while
the impact parameters of the scattering centers are denoted by $\vec s_j$.
 }}}
\end{figure}

Assuming that 
the scatterings on different nucleons are independent, the phase shifts
from each scattering simply add up,
\beq
\chi(\vec b,\vec r_1,\vec r_2,\dots\vec r_a)=
\sum_{j=1}^A\chi_j(\vec b-\vec s_j).
\eeq
Here, $\vec b$ is the impact parameter of the projectile and $\vec s_j$,
$j=1\dots A$ are the impact parameters of the $A$ nucleons in the nucleus,
see Fig.~\ref{fig:mscat}.
The amplitude for scattering off a nuclear target then can be written as
\beqn
F_{fi}^A&=&\frac{\imag p}{2\pi}\int d^2b\euler^{\imag\vec q\cdot\vec b}
\la f|1-\prod_{j=1}^A\euler^{\imag\chi_j(\vec b-\vec s_j)}|i\ra\\
\label{eq:nucf} 
&=&\frac{\imag p}{2\pi}\int d^2b\euler^{\imag\vec q\cdot\vec b}
\la f|1-\prod_{j=1}^A[1-\gamma_j(\vec b-\vec s_j)]|i\ra,
\eeqn
where $|f\ra$ and $|i\ra$ are the final and initial state of the target, 
respectively. In the second step, we introduced the profile function 
$\gamma(\vec b)$, which is related to the single-scattering amplitude 
$f(\vec q)$ by
\beq
\gamma(\vec b)=\frac{1}{2\pi\imag p}\int d^2q 
\euler^{-\imag\vec q\cdot\vec b}f(\vec q).
\eeq
Thus, we have expressed the nuclear scattering amplitude in terms of
the amplitude for scattering off a single nucleon.
 
In the case of a purely imaginary $f(\vec q)$, $\gamma(\vec b)$ 
is the probability of absorption of the projectile by a nucleon and 
the nuclear scattering amplitude, Eq.~(\ref{eq:nucf}) 
has a simple probabilistic interpretation. Namely, 
$1-\gamma_j(\vec b-\vec s_j)$
is the probability of not being absorbed by nucleon number $j$. 
Taking the product over all $j\in\{1\dots A\}$ yields the probability of not
being absorbed by any nucleon in the target. Finally, 
$1-\prod_{j=1}^A[1-\gamma_j(\vec b-\vec s_j)]$
is the probability that the projectile is absorbed by any of the nucleons.

Also, if one in addition
assumes that all nucleons in the target are identical, the nuclear cross 
section can be expressed in terms of the cross section for scattering on
a single nucleon,
\beqn\label{eq:g0}
\sigma_{tot}^A&=&\frac{4\pi}{p}{\rm Im} F_{ii}^A(t=0)\\&=&\label{eq:g1}
2\int d^2b\left(1-\left(1-\frac{\sigma_{tot}^NT(\vec b)}{2A}\right)^A
\right)\\
\label{eq:g2}&\approx&
2\int d^2b\left(1-\exp\left(-\frac{\sigma_{tot}^NT_A(\vec b)}{2}\right)\right),
\eeqn
where the nuclear thickness function $T_A(\vec b)$ is the integral over the
nuclear density,
\beq
T_A(\vec b)=\int_{-\infty}^\infty dz \rho_A(\vec b,z).
\eeq
The simple expression, Eq.~(\ref{eq:g1}), resums all multiple scattering
terms.
We stress that the probabilistic interpretation of Eq.~(\ref{eq:nucf}) as well
as Eqs.~(\ref{eq:g1}) and (\ref{eq:g2}) only hold for a purely imaginary
$f(\vec q)$.

The meaning of the nuclear scattering amplitude, Eq.~(\ref{eq:nucf}), is
further explained by expanding the probability of particle absorption
by any of the nucleons in powers of $\gamma(\vec b)$,
\beqn
\Gamma(\vec b,\vec r_1,\vec r_2,\dots\vec r_a)&\equiv&
1-\prod_{j=1}^A[1-\gamma_j(\vec b-\vec s_j)]\\
\label{eq:double}
&=&\sum_{k=1}^A\gamma_k(\vec b-\vec s_k)
-\sum_{k>j}\gamma_k(\vec b-\vec s_k)\gamma_j(\vec b-\vec s_j)+\dots\, .
\eeqn
\begin{figure}[t]
\centerline
{
  \scalebox{0.6}{\includegraphics{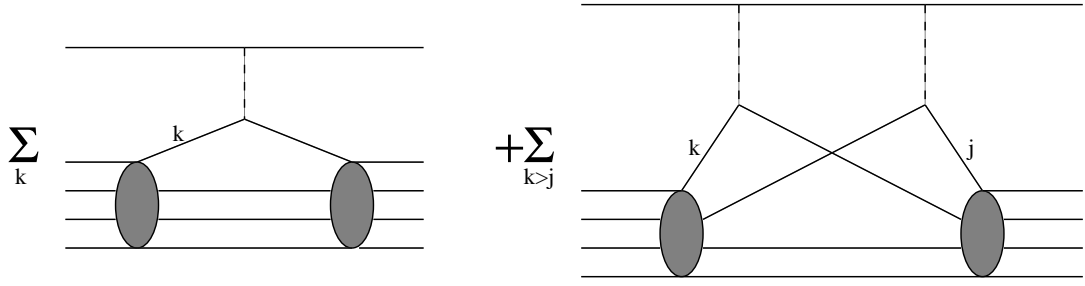}}
 }
{\parbox[b]{13cm}{
\caption{\label{fig:double}\em Illustration of the single and double 
scattering terms in Eq.~(\ref{eq:double}). The coherent sum over all
graphs leads to interferences that reduce the total cross section.
 }}}
\end{figure}
The first two terms in this expansion are illustrated in 
Fig.~\ref{fig:double}. The first term in Eq.~(\ref{eq:double}) is just the sum
of single scattering amplitudes. However, different nucleons in the nucleus
compete to interact with the projectile. 
This effect is contained in the second term 
in Eq.~(\ref{eq:double}), which reduces the cross section.
This reduction is an interference effect that appears because
the amplitudes for scattering on different nucleons have to be added 
coherently. This destructive interference can be observed in experiment as
shadowing in hadron-nucleus interactions (eclipse effect in deuterium).
Note, however, that shadowing is not completely explained by Glauber theory, 
as will be explained in the following section.

The easiest application of Glauber multiple scattering theory to nuclear
systems is the calculation of the inelastic nucleus-nucleus ($AB$) 
cross section, which can be
written as
\beq
\sigma_{AB}^{in}=\int d^2b(1-P_0(b)). \label{sigma3}
\eeq
here,  $P_0(b)$ is the probability that no interaction takes place,
\beq
P_0(b)=(1-\sigma_{NN}^{in}T_{AB}(b))^{AB},
\eeq
where the nuclear overlap function is given by
\beq
T_{AB}(\vec b)=\int d^2s T_A(\vec s)T_B(\vec b-\vec s).
\eeq
(Obviously, $1-P_0(b)$ is then the probability of an inelastic interaction, 
and the meaning of Eq. (\ref{sigma3}) becomes very transparent.)
As it is common, we have labeled the two nuclei by their atomic mass numbers
$A$ and $B$.

Another application is the calculation of inclusive particle spectra. With the 
help of crossing symmetry, the cross section for production of a particle of
type $a$ in an $AB$ collision, $AB\to aX$, can be calculated from the
total cross section of the process $\bar aAB\to X$, where $\bar a$ is the 
antiparticle of $a$. According to so-called AGK cutting rules \cite{agk}, the
nuclear cross section for this process is given by
\beq
E\frac{d^3\sigma^a_{AB}}{d^2bd^3p}=T_{AB}(\vec b)E\frac{d^3\sigma^a_{NN}}{d^3p}.
\eeq
Integration over impact parameter $b$ yields
\beq
E\frac{d^3\sigma^a_{AB}}{d^3p}=AB\;E\frac{d^3\sigma^a_{NN}}{d^3p},
\eeq
and correspondingly the charged particle multiplicity would scale proportional
to $AB$,
\beq\label{eq:nch}
\frac{dn_{ch}}{d\eta}=AB\;\frac{1}{\sigma^{in}_{AB}}
\frac{d\sigma^{ch}_{NN}}{d\eta},
\eeq
the meaning of which is obvious -- if collisions are truly independent, 
the resulting multiplicity should scale with the number of collisions, $AB$. 

However, the relation (\ref{eq:nch}) appears to be badly violated in experiment.
What went wrong? 
It appears that the disagreement between the result 
Eq.~(\ref{eq:nch}) and experimental data is due to the fact that there
are important corrections to the Glauber multiple scattering theory, which 
we neglected so far. These corrections are known as Gribov's inelastic
shadowing \cite{glaubergribov} and will be the subject of the next section.

\subsubsection{Gribov's ``Inelastic Shadowing''}

\begin{figure}[t]
\centerline
{
  \scalebox{0.7}{\includegraphics{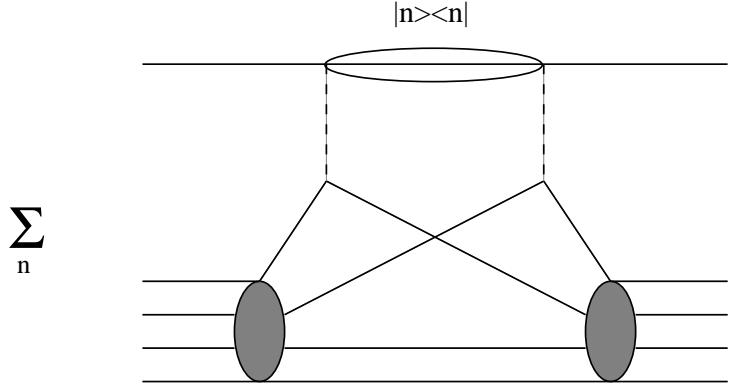}}
 }
{\parbox[b]{13cm}{
\caption{\label{fig:diff}\em If the projectile is a composite particle, it
can be excited by the interaction. Therefore, this graph will interfere
with the double scattering graph in Fig.~\ref{fig:double}.
 }}}
\end{figure}

The assumed independence of nucleon--nucleon collisions is violated by the
diagrams of the type of 
Fig.~\ref{fig:diff}, where the projectile is excited into a state $|n\ra$
by the interaction. The diagram in Fig.~\ref{fig:diff} does not describe 
independent collisions, and at high energies it will
interfere with the double scattering graph in Fig.~\ref{fig:double}.

The excitation of an inelastic state in the scattering 
is accompanied by a longitudinal momentum transfer
\beq\label{eq:dp}
\Delta p_L=\frac{M_f^2-M_i^2}{2p},
\eeq
where $M_f$ is the invariant mass of the excited system and $M_i$ is the 
invariant mass of the projectile in the initial state. The diagram in
Fig.~\ref{fig:diff} is only important if it can make a significant
contribution to the forward scattering amplitude $F^A_{ii}$. This requires
that the longitudinal momentum transfer must be 
so small that the nucleus has a 
chance to remain intact, {\em i.e.}
\beq\label{eq:cond}
\Delta p_L R_A\lsim 1,
\eeq
where $R_A$ is the nuclear radius. Apparently, this condition is fulfilled
for sufficiently large values of the projectile momentum $p$ in 
Eq.~(\ref{eq:dp}). Thus, as it was first found by Gribov in 
\cite{glaubergribov},
Glauber theory receives important corrections at high energy.

The condition, Eq.~(\ref{eq:cond}), which 
determines, whether Gribov's inelastic
shadowing becomes relevant, leads us to the important quantum mechanical
concept of 
{\em formation time}, or {\em formation length}. 
The formation time is the lifetime of the excitation $|n\ra$ in
Fig.~\ref{fig:diff} in the target rest frame and the formation length
is the longitudinal distance over which the excited state $|n\ra$ lives.
At high energy, of course, both quantities are identical. The formation 
time/length can be determined in a time-dependent and in a time-independent
approach. 

In the time dependent formulation, one starts from the energy-time
uncertainty relation,
\beq
\Delta E\Delta t\gsim 1.
\eeq
The lifetime of the excitation in rest frame of the projectile is given by
\beq
\tau_f\approx\frac{1}{M_f-M_i}.
\eeq
In order to obtain the formation time, we have to transform $\tau_f$ to
the target rest frame, by multiplying $\tau_f$
with the relativistic $\gamma$-factor,
\beq
t_f\equiv\Delta t=\gamma\tau_f=\frac{p}{\bar M}\tau_f,
\eeq
where
\beq
\bar M=\frac{M_f+M_i}{2}.
\eeq
We finally obtain for the formation time
\beq\label{eq:tf}
t_f\approx\frac{2p}{M_f^2-M_i^2}.
\eeq

In the time-independent approach, one starts from the coordinate-momentum
uncertainty relation,
\beq\label{eq:dpdx}
\Delta p_L\Delta z\gsim 1.
\eeq
The longitudinal momentum transfer was already given in Eq.~\ref{eq:dp}. 
It is calculated in the following way,
\beq
\Delta p_L=\sqrt{E^2-M_i^2}-\sqrt{E^2-M_f^2}
\approx\frac{M_f^2-M_i^2}{2p}.
\eeq
According to the uncertainty relation Eq.~(\ref{eq:dpdx}), the excited
state lives over the longitudinal extension
\beq\label{eq:lf}
l_f\equiv\Delta z\approx\frac{1}{\Delta p_L}\approx\frac{2p}{M_f^2-M_i^2}.
\eeq
As expected, the formation length is identical to the formation time given in
Eq.~(\ref{eq:tf}).

We see from Eqs.~(\ref{eq:tf}) and (\ref{eq:lf}) that for large initial
projectile momentum, the process develops at large longitudinal 
distances in the target rest frame. At the high center of mass energies of 
RHIC and LHC, the coherence length will be much larger than the nuclear radius
and all scattering processes will be governed by coherence effects (the coherence 
length becomes as long as several hundreds fm).

\subsection{Elementary hadron--hadron scattering at high energies} 

All of the formalism presented above is completely independent of the underlying
interaction. Before concluding this section, we will briefly discuss the main 
properties of hadron--hadron scattering at high energies.
Let us begin by 
listing some empirical facts
about hadronic cross sections: 

\begin{itemize}
\item{Total hadronic cross sections are approximately constant at {\em cm} 
energies of order $\sqrt{s}\sim 20\GeV$ and slowly rise, 
$\sigma_{tot}\sim s^{0.08}$, up to the highest energies accessible in 
experiment
(Tevatron energy, $\sqrt{s}=1.8$ TeV).}
\item{The diffraction cone shrinks as energy increases, indicating that the
size of the hadron increases with energy.}
\item{The mean transverse momentum of produced particles is approximately
constant or increases only slowly with energy, respectively.}  
\end{itemize}

\begin{figure}[t]
\centerline
{
  \scalebox{0.9}{\includegraphics{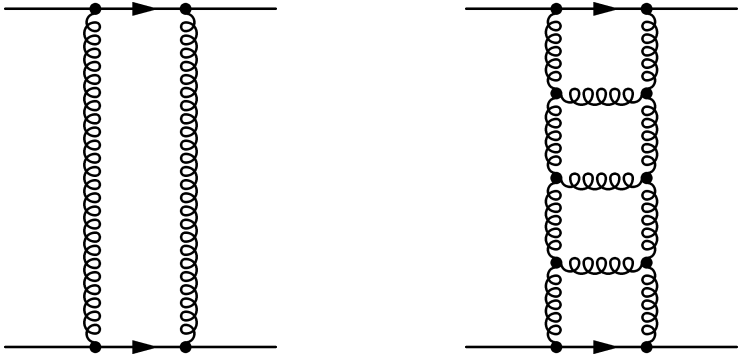}}
 }
{\parbox[b]{13cm}{
\caption{\label{fig:ladder}\em Double gluon exchange (left), yields an
imaginary scattering amplitude and a constant
cross section. The rise of hadronic cross sections and the shrinkage
of the diffraction cone at high energy is due to
radiation of additional gluons (right).
 }}}
\end{figure}

The (approximate) constancy of the total cross section in the framework of QCD 
implies that high
energy hadronic scattering is dominated by two gluon exchange \cite{low}, see 
Fig.~\ref{fig:ladder} (left). The two gluon exchange model also yields a 
purely imaginary forward scattering amplitude. In order to explain the 
increase of the total cross section with energy, one has to take the radiation
of additional gluons into account, see Fig.~\ref{fig:ladder} (right).
The probability of gluon emission is proportional to 
$\alpha_sy\sim\alpha_s\ln s$, where $y$ is rapidity. 
Thus, each gluon radiation in 
Fig.~\ref{fig:ladder} (right) contributes a factor $\ln s$ to the total cross
section.
Resumming an infinite number of gluon emissions ordered in rapidity,
one finds that the total cross section behaves like
\beq
\sigma_{tot}\propto\sum_n\frac{(\ln s)^n}{n!}a^n=s^a,
\eeq
where $a\propto\alpha_s$.

This gluon radiation also explains the shrinkage of the diffraction cone.
At high energy, the $t$-differential cross section in hadronic collisions
behaves like
\beq
\frac{d\sigma}{dt}=\left.\frac{d\sigma}{dt}\right|_{t=0}\euler^{-B(s)|t|},
\eeq
where $B(s)\propto\ln s$ increases with energy.
Such a behavior emerges, if the elastic scattering amplitude in impact 
parameter space is given by
\beq
f_{el}\propto\exp\left(\frac{b}{R^2_h(s)}\right),
\eeq
where the effective hadron radius $R^2_h(s)$ increases as a function of energy.
Therefore, the shrinkage of the diffraction peak suggests an
increase of hadronic sizes with energy. 
In QCD, this can be understood in the following way: Gluons are radiated
off the projectile with different transverse momenta. As rapidity, or energy, 
increases, these gluons perform a random walk in the impact 
parameter plane and correspondingly, the transverse size of the gluon cloud 
surrounding the projectile increases. This can be regarded as a diffusion
process in the impact parameter plane, in which rapidity plays the role of time.

The slow increase of the mean transverse momentum with energy is 
likely to be related to asymptotic freedom. Indeed, at large transverse
momentum, $p_\perp\gg\Lambda_{QCD}$, the strong coupling constant becomes
small, $\alpha_s(p_\perp)\ll1$, which suppresses the production of high
$p_\perp$ particles.
 
\begin{figure}[t]
\centerline
{
  \scalebox{0.95}{\includegraphics{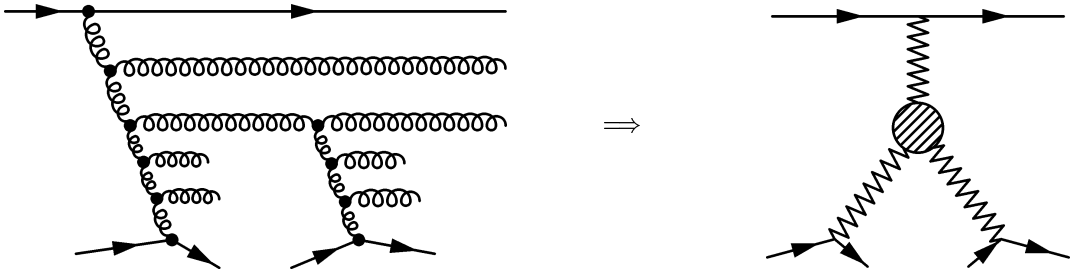}}
 }
{\parbox[b]{13cm}{
\caption{\label{fig:fusion}\em At sufficiently high energy, emitted gluons
can themselves develop showers. In the squared amplitude (right), the gluons 
combine to ladders, which are denoted by zigzag lines. Fusion of gluon ladders
is the mechanism behind gluon shadowing. A resummation of fan
diagrams like in the left figure, corresponds to classical solutions \cite{schwimmer} 
of Reggeon Field Theory \cite{reggeonfield}. 
 }}}
\end{figure}

Eventually, the power-like growth of the total hadronic cross section will
violate the Froissart-Martin bound \cite{froissart}, 
which states that as a consequence of
unitarity and analyticity, total cross sections cannot rise faster than
\beq
\sigma_{tot}\lsim C\ln^2s,
\eeq
where $C$ is a constant.  
At sufficiently high energy, emitted gluons can develop
showers themselves, see Fig.~\ref{fig:fusion}. Due to this process, the
projectile sees a reduced gluon density in the target and the 
growth of the cross section is slowed down. This effect, in the squared amplitude, 
realizes a QCD realization of Gribov's inelastic shadowing (see Fig.~\ref{fig:fusion}.)
 Even though such unitarity corrections 
might already be present in proton-antiproton
scattering at Tevatron, they will be much more pronounced in nuclear collisions
at RHIC. 

As the magnitude of these effects increases with energy and/or atomic number of the 
colliding nuclei, the classification of diagrams in terms of individual nucleon--nucleon 
amplitudes (or parton ladders) rapidly starts to lose sense -- the non--linear effects 
become extremely important. The treatment of nuclear interactions in this high--density 
regime will be considered in the following section.

\section{Classical Chromodynamics of Relativistic Heavy Ion 
Collisions}

\subsection{QCD in the classical regime}
 
\begin{figure}[t]
\centerline
{
  \scalebox{0.5}{\includegraphics{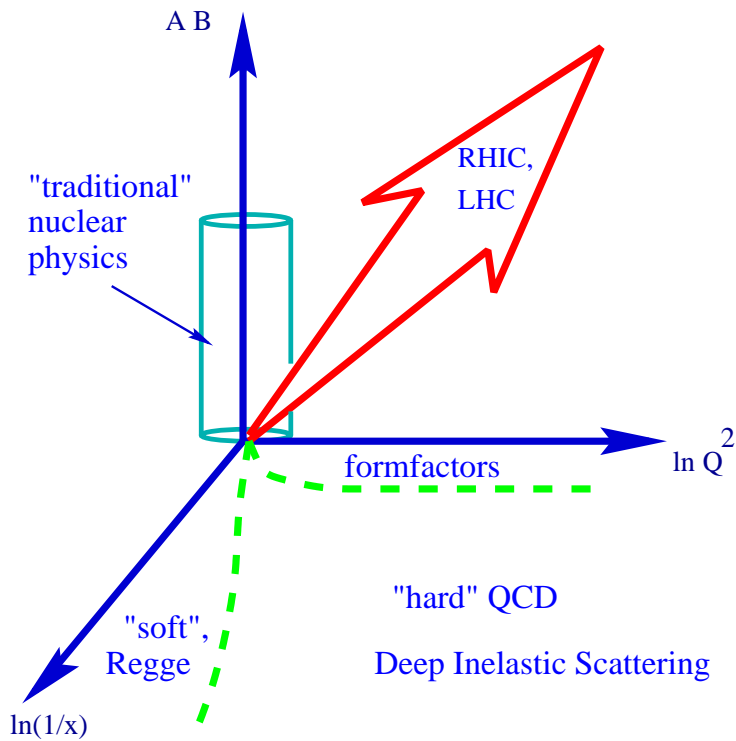}}
 }
{\parbox[b]{13cm}{
\caption{\label{fig:rhicdiag}\em
The place of relativistic heavy ion physics in the study of QCD;
the vertical axis is the product of atomic numbers of projectile and target,
and the horizontal axes are the momentum transfer $Q^2$ and rapidity
$y=\ln(1/x)$ ($x$ is the Bjorken scaling variable).
 }}}
\end{figure}

Most of the applications of QCD so far have been  
limited to the short distance regime of high momentum transfer, 
where the theory becomes weakly coupled and can be linearized.
While this is the only domain where our theoretical tools based 
on perturbation theory are adequate, this is also the domain in 
which the beautiful non--linear structure of QCD does not yet reveal 
itself fully. On the other hand, as soon as we decrease the momentum 
transfer in a process, the dynamics rapidly becomes non--linear, but our  
understanding is hindered by the large coupling. 
Being perplexed by this problem, one is 
tempted to dream about an environment in which the coupling is weak, 
allowing a systematic theoretical treatment, but the fields are strong, 
revealing the full non--linear nature of QCD. 
We are going to argue now that this environment can be created on Earth 
with the help of relativistic heavy ion colliders.    
Relativistic heavy ion collisions allow to probe QCD in the non--linear 
regime of high parton density and high color field strength, 
see Fig.~\ref{fig:rhicdiag}.

It has been conjectured long time ago that the dynamics of QCD  
in the high density domain  may become qualitatively different: in parton language, 
this is best described in terms of {\it parton saturation} \cite{GLR,MQ,BM}, and in the language 
of color fields -- in terms of the {\it classical} Chromo--Dynamics \cite{MV}; see the lectures 
\cite{ILM} and \cite{AHM} and references therein. 
In this high density regime,  
the transition amplitudes are dominated not by quantum fluctuations, but by 
the configurations of classical field containing large, $\sim 1/\alpha_s$, 
numbers of gluons. One thus uncovers new 
non--linear features of QCD, 
which cannot be investigated in the more traditional applications
based on the perturbative approach. 
The classical color fields in the initial nuclei (the 
``color glass condensate'' \cite{ILM}) can be thought of as 
either perturbatively generated, or 
as being a topologically non--trivial superposition of the Weizs{\"a}cker-Williams 
radiation and the quasi--classical vacuum fields \cite{inst,inst1,KKL}.    

\subsubsection{Geometrical arguments}
 
Let us consider an external probe $J$ interacting with the 
nuclear target of atomic number $A$. At small values of Bjorken $x$, 
by uncertainty principle the interaction develops over large 
longitudinal distances $z \sim 1/mx$, where $m$ is the 
nucleon mass. As soon as $z$ becomes larger than the nuclear diameter, 
the probe cannot distinguish between the nucleons located on the front and back edges 
of the nucleus, and all partons within the transverse area $\sim 1/Q^2$ 
determined 
by the momentum transfer $Q$ participate in the interaction coherently. 
The density of partons in the transverse plane is given by
\beq
\rho_A \simeq {x G_A(x,Q^2) \over \pi R_A^2} \sim A^{1/3},
\eeq
where we have assumed that the nuclear gluon distribution  
scales with the number of nucleons $A$. The probe interacts with 
partons with cross section $\sigma \sim \alpha_s / Q^2$; therefore, 
depending on the magnitude of momentum transfer $Q$, atomic number $A$, 
and the value of Bjorken $x$, one may encounter two regimes:
\begin{itemize}
\item{$\sigma \rho_A \ll 1$ -- this is a familiar ``dilute'' regime of 
incoherent interactions, which is well described by the methods of 
perturbative QCD;}
\item{$\sigma \rho_A \gg 1$ -- in this regime, we deal with a dense 
parton system. Not only do the ``leading twist'' expressions become 
inadequate, but also the expansion in higher twists, i.e. in 
multi--parton correlations, breaks down here.}
\end{itemize} 

\begin{figure}[h]
\begin{minipage}[t]{140mm}
\includegraphics[width=14pc]{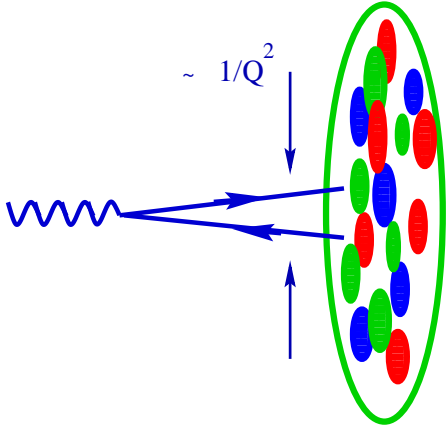}
\caption{{\em Hard probe interacting with the nuclear target 
resolves the transverse distance $\sim 1/\sqrt{Q}$ ($Q^2$ is the square of the 
momentum transfer) and, in the target rest frame, the longitudinal 
distance $\sim 1/(m x)$ ($m$ is the nucleon mass and $x$ 
the Bjorken variable).}} 

\label{fig:satpict}
\end{minipage}
\end{figure}

The border between the two regimes can be found from the condition 
$\sigma \rho_A \simeq 1$; it determines the critical value of the 
momentum transfer (``saturation scale''\cite{GLR}) at which the parton system 
becomes to look dense to the probe\footnote{Note that since 
$Q_s^2 \sim A^{1/3}$, 
which is the length of the target, this expression 
in the target rest frame can also be understood as describing a broadening 
of the transverse momentum resulting from the multiple re-scattering 
of the probe.}:
\beq 
Q_s^2 \sim  \alpha_s \ {x G_A(x,Q_s^2) \over \pi R_A^2}. \label{qsat}
\eeq
In this regime, the number of gluons from (\ref{qsat}) is given by 
\beq
x G_A(x,Q_s^2) \sim {\pi \over \alpha_s(Q_s^2)}\ Q_s^2 R_A^2, \label{glustr}
\eeq
where $Q_s^2 R_A^2 \sim A$. 
One can see that the number of gluons 
is proportional to the {\em inverse} of $\alpha_s(Q_s^2)$, and 
becomes large in the weak coupling regime. In this regime,
as we shall now discuss, the dynamics is likely to become 
essentially classical. 

\subsubsection{Saturation as the classical limit of QCD}

Indeed, the condition (\ref{qsat}) can be derived in the following, 
rather general, way. As a first step, let us note that 
the dependence of the action corresponding to the 
Lagrangian (\ref{eq:lag}) on the coupling constant is given by  
\beq
S \sim \int {1 \over g^2}\ G_{\mu \nu}^a  G_{\mu \nu}^a 
\ d^4 x. \label{act}
\eeq
Let us now consider a classical configuration of gluon fields; by definition, 
$G_{\mu \nu}^a$ in such a configuration does not depend on 
the coupling, and the action is large, $S \gg \hbar$. The number of 
quanta in such a configuration is then
\beq
N_g \sim {S \over \hbar} \sim {1 \over \hbar \ g^2}\ \rho_4 V_4, \label{numb}
\eeq
where we re-wrote (\ref{act}) as a product of four--dimensional 
action density $\rho_4$ and the four--dimensional volume $V_4$. 
 
Note that since (\ref{numb}) depends only on the product of the Planck constant $\hbar$ and 
the coupling $g^2$, the classical limit $\hbar \to 0$ is indistinguishable from the 
weak coupling limit $g^2 \to 0$. The weak coupling limit of small $g^2 = 4 \pi \alpha_s$ 
therefore corresponds to the semi--classical regime.

The effects of non--linear interactions among the gluons become 
important when $\partial_{\mu} A_{\mu} \sim A_{\mu}^2$ 
(this condition can be made explicitly gauge invariant if we derive it 
from the expansion of a correlation function of gauge-invariant 
gluon operators, e.g., $G^2$). In momentum space, this 
equality corresponds to 
\beq
Q_s^2 \sim (A_{\mu})^2 \sim (G^2)^{1/2} = 
\sqrt{\rho_4}; \label{nonlin}
\eeq
$Q_s$ is the typical value of the gluon momentum below which 
the interactions become essentially non--linear. 

Consider now a nucleus $A$ boosted to a high momentum. By uncertainty 
principle, the gluons with transverse momentum $Q_s$ are extended 
in the longitudinal and proper time directions by $\sim 1/Q_s$; 
since the transverse area is $\pi R_A^2$, the four--volume 
is $V_4 \sim \pi R_A^2 / Q_s^2$. The resulting four--density from 
(\ref{numb}) is then 
\beq
\rho_4 \sim \alpha_s\ {N_g \over V_4} \sim \alpha_s\ {N_g\ Q_s^2 
\over \pi R_A^2} 
\sim Q_s^4, \label{class}
\eeq
where at the last stage we have used the non--linearity condition (\ref{nonlin}),  
$\rho_4 \sim Q_s^4$. It is easy to see that (\ref{class}) coincides with the 
saturation condition (\ref{qsat}), since the number of gluons in the 
infinite momentum frame $N_g \sim x G(x,Q_s^2)$. 
\vskip0.3cm

In view of the significance of saturation criterion for the rest of the material 
in these lectures, let us present yet another argument, traditionally followed 
in the discussion of classical limit in electrodynamics \cite{LL}.
The energy of the gluon field per unit volume is $\sim \vec{E}^{a 2}$. The number 
of elementary ``oscillators of the field'', also per unit volume, is $\sim \omega^3$.
To get the number of the quanta in the field we have to divide the energy of the field 
by the product of the number of the oscillators $\sim \omega^3$ and the average energy 
$\hbar \omega$ of the gluon:
\be
N_{{\vec k}} \sim {\vec{E}^{a 2} \over \hbar \omega^4}. \label{qc3}
\ee

The classical approximation holds when $N_{{\vec k}} \gg 1$. Since the energy $\omega$ of the 
oscillators is related to the time $\Delta t$ over which the average energy is computed 
by $\omega \sim 1/\Delta t$, we get 
\be
\vec{E}^{a 2} \gg {\hbar \over (\Delta t)^4}. \label{llcon}  
\ee
Note that the quantum mechanical uncertainty principle for the energy of the field 
reads 
\be
\vec{E}^{a 2} \  \omega^4 \sim \hbar,
\ee
so the condition (\ref{llcon}) indeed defines the quasi--classical limit.

Since $\vec{E}^{a 2}$ is proportional to the action density $\rho_4$, and the typical time 
is $\Delta t \sim 1/k_{\perp}$, using (\ref{class}) we 
finally get that the classical description applies when
\beq 
k_{\perp}^2 < \alpha_s {N_g \over \pi R_A^2} \equiv Q_s^2.
\eeq

\subsubsection{The absence of mini--jet correlations}

When the occupation numbers of the field become large, the matrix elements of the creation and annihilation 
operators of the gluon field defined by
\beq
\hat{A}^{\mu} = \sum_{\vec{k}, \alpha} (\hat{c}_{\vec{k} \alpha} A^{\mu}_{\vec{k} \alpha} + 
\hat{c}^\dagger_{\vec{k} \alpha} A^{\mu *}_{\vec{k} \alpha})
\eeq
become very large,
\beq
N_{\vec{k} \alpha} = \langle  \hat{c}^\dagger_{\vec{k} \alpha} \hat{c}_{\vec{k} \alpha} \rangle \gg 1,
\eeq
so that one can neglect the unity on the r.h.s. of the commutation relation
\beq
\hat{c}_{\vec{k} \alpha}   \hat{c}^\dagger_{\vec{k} \alpha} - 
\hat{c}^\dagger_{\vec{k} \alpha} \hat{c}_{\vec{k} \alpha} = 1
\eeq 
and treat these operators as classical $c-$numbers.
 
This observation, often used in condensed matter physics, especially in the theoretical 
treatment of superfluidity, has important consequences for gluon production -- in particular, it implies that 
the correlations among the gluons in the saturation region can be neglected:
\beq
\langle A(k_1) A(k_2) ... A(k_n) \rangle \simeq \langle A(k_1) \rangle  \langle A(k_2) \rangle ... 
\langle A(k_n) \rangle. \label{abscor}
\eeq
Thus, in contrast to the perturbative picture, where the produced mini-jets have strong back-to-back 
correlations, the gluons resulting from the decay of the 
classical saturated field are uncorrelated at $k_{\perp} \lsim Q_s$.

Note that the amplitude with the factorization property (\ref{abscor}) is called 
point--like. However, the relation (\ref{abscor}) cannot be exact if we consider 
the correlations of final--state hadrons -- the gluon mini--jets cannot transform 
into hadrons independently. These correlations caused by color confinement however 
affect mainly hadrons with close three--momenta, as opposed to the perturbative 
correlations among mini--jets with the opposite three--momenta.

It will be interesting to explore the consequences of the factorization 
property of the classical gluon field (\ref{abscor}) for the HBT correlations of 
final--state hadrons. It is likely that the HBT radii in this case reflect  
the universal color correlations in the hadronization process.  

\vskip0.3cm

Another interesting property of classical fields follows from the relation 
\beq
\langle  (\hat{c}^\dagger_{\vec{k} \alpha} \hat{c}_{\vec{k} \alpha})^2 \rangle -
\langle  \hat{c}^\dagger_{\vec{k} \alpha} \hat{c}_{\vec{k} \alpha} \rangle^2 = 
 \langle  \hat{c}^\dagger_{\vec{k} \alpha} \hat{c}_{\vec{k} \alpha} \rangle, \label{fluctc}
\eeq
which determines the fluctuations in the number of produced gluons. 
We will discuss the implications of Eq. (\ref{fluctc}) for the multiplicity fluctuations 
in heavy ion collisions later.

\subsection{Classical QCD in action}

\subsubsection{Centrality dependence of hadron production}

In nuclear collisions, the saturation scale becomes a function of centrality; 
a generic feature of the quasi--classical 
approach -- the proportionality of the number of gluons to the inverse 
of the coupling constant (\ref{numb}) -- thus leads to definite predictions \cite{KN} 
on the centrality dependence of multiplicity. 

Let us first present the argument on a qualitative level.
At different centralities 
(determined by the impact parameter of the collision), the average density of partons 
(in the transverse plane)  
participating in the collision is very different. This density $\rho$ 
is proportional to the average length of nuclear material involved in the collision, 
which in turn approximately scales with the power of the number $N_{part}$ 
of participating nucleons, $\rho \sim N_{part}^{1/3}$.
The density of partons defines the value of the saturation scale, and so we expect
\be
Q_s^2 \sim N_{part}^{1/3}.
\ee
The gluon multiplicity is then, as we discussed above, is
\be
{dN_g \over d \eta} \sim {S_A \ Q_s^2 \over \alpha_s(Q_s^2)}, \label{mults}
\ee
where $S_A$ is the nuclear overlap area, 
determined by atomic number and the centrality of collision. Since $S_A\ Q_s^2 \sim N_{part}$ 
by definitions of the transverse density and area, from (\ref{mults}) we get
\be
{dN_g \over d \eta} \sim N_{part}\ \ln N_{part}, \label{mults1}
\ee
which shows that the gluon multiplicity shows a logarithmic deviation from the scaling 
in the number of participants.

To quantify the argument, we need to explicitly evaluate the average density of 
partons at a given centrality. This can be done by using Glauber theory, which 
allows to evaluate the differential cross section of the nucleus--nucleus interactions. 
The shape of the multiplicity distribution at a given (pseudo)rapidity $\eta$ can 
then be readily obtained by using the formulae introduced in section 2:
\be
\frac {d \sigma} {d n} = \int d^2b \ {\cal P}(n;b)\ (1 - P_0(b)),  
\ee
where $P_0(b)$ is the probability of no interaction among the nuclei at a given 
impact parameter $b$: 
\be
P_0(b) = (1 - \sigma_{NN} T_{AB}(b))^{AB}; 
\ee
$\sigma_{NN}$ is the inelastic nucleon--nucleon cross section, and $T_{AB}(b)$ is the 
nuclear overlap function for the collision of nuclei with atomic numbers A and B; 
we have used the three--parameter Woods--Saxon nuclear density distributions \cite{tables}.

The correlation function ${\cal P}(n;b)$ is given by
\be
{\cal P}(n;b) = \frac{1}{\sqrt{2\pi a \bar{n}(b)}}\ \exp\left( - \frac{(n - \bar{n}(b))^2}{2 a 
\bar{n}(b)}\right),
\ee
here $\bar{n}(b)$ is the mean multiplicity at a given impact parameter $b$; 
the formulae for the number of participants and the number of binary 
collisions can be found in \cite{KLNS}.  The parameter 
$a$ describes the strength of fluctuations; for the classical gluon field, as follows from (\ref{fluctc}), 
$a = 1$. However, the strength of fluctuations can be changed by the subsequent evolution of the system 
and by hadronization process. Moreover, in a real experiment, the strength of fluctuations strongly 
depends on the acceptance. In describing the PHOBOS distribution \cite{PHOBOS}, we have found that the value $a = 0.6$ 
fits the data well. 
 
\begin{figure}[htbp] 
\epsfig{file=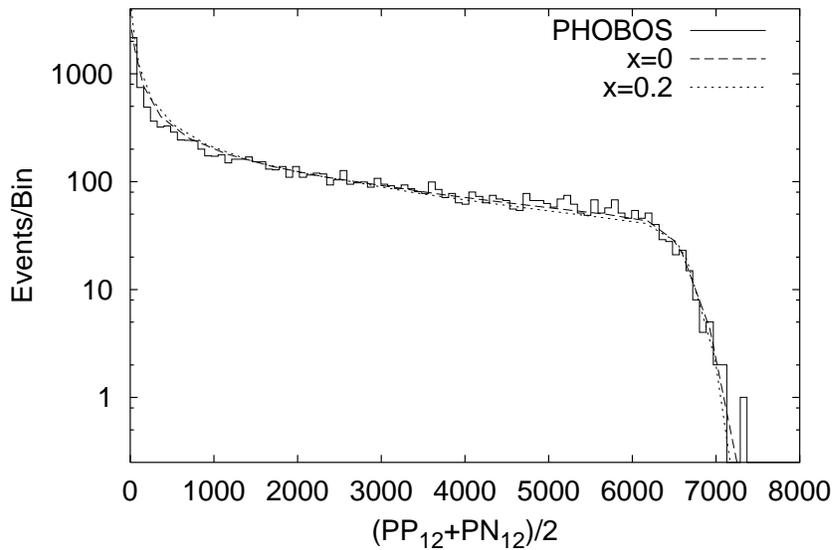,width=11cm}
\caption{\em Charged multiplicity distribution at $\sqrt{s}=130$ A GeV; solid line (histogram) -- 
PHOBOS result; dashed line -- distribution corresponding to 
participant scaling ($x=0$); 
dotted line -- distribution corresponding to the $37 \%$ admixture of ``hard'' component 
in the multiplicity; see 
text for details.}
\label{fig1321}
\end{figure}

\begin{figure}[htbp]
\epsfig{file=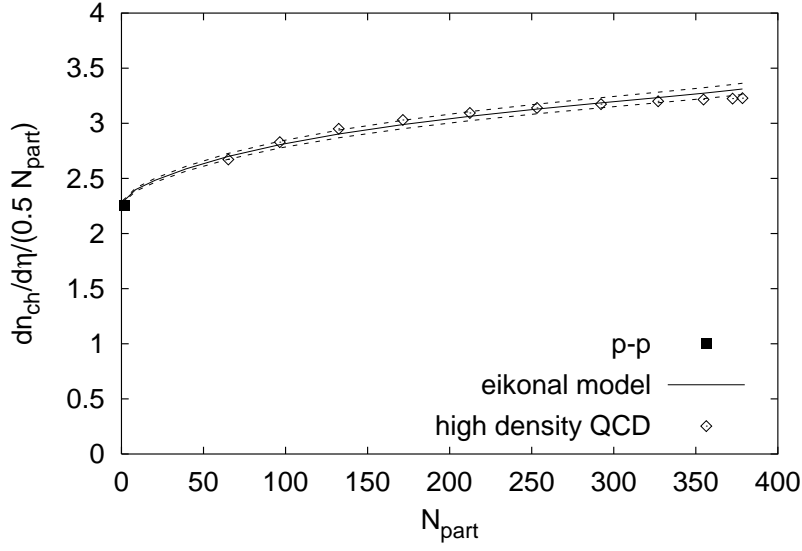,width=11cm}
\caption{\em
Centrality dependence of the charged multiplicity per participant pair near $\eta = 0$ at 
$\sqrt{s} = 130$ A GeV; the curves represent the prediction based on the conventional eikonal 
approach, while the diamonds correspond to the high density QCD prediction (see text). The square 
indicates the $pp$ multiplicity.}
\label{fig:centr}
\end{figure}

In Fig. \ref{fig1321}, we compare the 
resulting distributions for two different assumptions about the scaling of multiplicity with the number 
of participants to the PHOBOS 
experimental distribution, measured in the interval $3 < |\eta| < 4.5$.  
One can see that almost independently of theoretical assumptions about the dynamics of multiparticle production, 
the data are described quite well. At first this may seem surprising; the reason for this result is that at high energies, 
heavy nuclei are almost completely ``black''; unitarity then implies that  the shape of the cross section 
is determined almost entirely by the nuclear geometry.   
We can thus use experimental differential cross sections as 
a reliable handle on centrality. This gives us a possibility to compute the dependence of the 
saturation scale on centrality of the collision, and thus to predict the centrality dependence of 
particle multiplicities, shown in Fig. \ref{fig:centr}.  
(see \cite{KN} for details).

\vskip0.1cm

\subsubsection{Energy dependence}

\begin{figure}[h]
\begin{turn}{-90}
{\epsfig{file=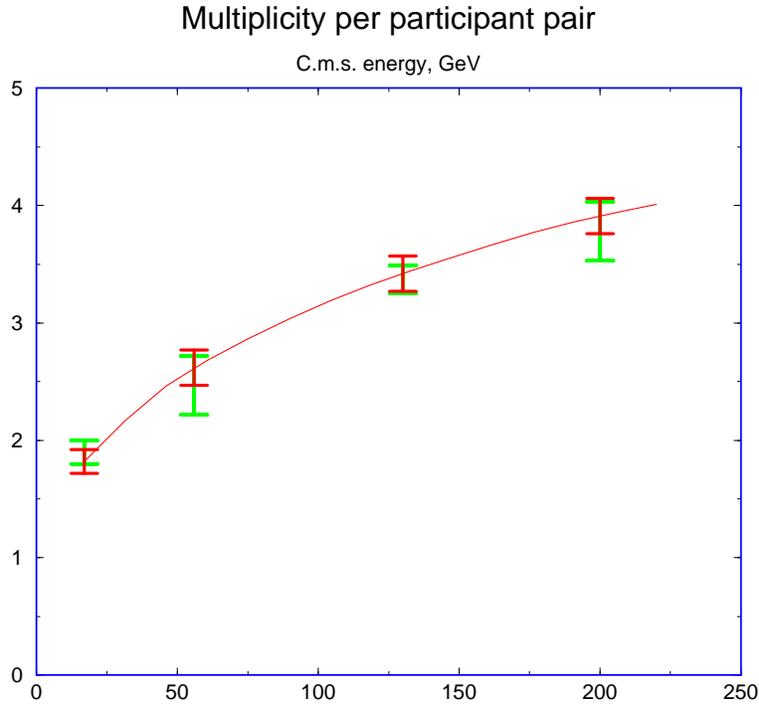,width=22pc}}
\end{turn}
\caption{\em
Energy dependence of charged multiplicity per participant pair at RHIC energies; solid line is the result 
(\ref{endep}).}
\label{fig:mult}
\end{figure}

Let us now turn to the discussion of energy dependence of hadron production. In semi--classical scenario, 
it is determined by the variation of saturation scale 
$Q_s$ with Bjorken $x = Q_s / \sqrt{s}$. This variation, in turn, is determined by the 
$x-$ dependence of the gluon structure function. 
In the saturation approach, the gluon distribution is related to the saturation scale 
by Eq.(\ref{qsat}). 
A good description of HERA data is obtained with saturation scale $Q_s^2 = 1 \div 2\ \rm{GeV}^2$
with $W$ - dependence ($W \equiv \sqrt{s}$ is the center-of-mass energy available 
in the photon--nucleon system)  \cite{GW} 
\beq
Q^2_s \,\,\propto\, W^{\lambda},
\eeq
where $\lambda \simeq 0.25 \div 0.3$. In spite of significant uncertainties in the determination 
of the gluon structure functions, perhaps even more important is the observation \cite{GW} that the 
HERA data exhibit scaling when plotted as a function of variable 
\beq
\tau \,=\, {Q^2 \over Q_0^2} \ \left({x \over x_0}\right)^{\lambda}, 
\eeq
where the value of $\lambda$ is again within the limits $\lambda \simeq 0.25 \div 0.3$. 
In high density QCD, this scaling is a consequence 
of the existence of dimensionful scale \cite{GLR,MV} 
\beq
Q_s^2(x) = Q_0^2 \ (x_0 / x)^{\lambda}. 
\eeq
Using the value of $Q_s^2 \simeq 2.05\ {\rm GeV}^2$ extracted \cite{KN} at $\sqrt{s} = 130$ GeV and $\lambda = 0.25$ 
\cite{GW} used in \cite{KL}, equation (\ref{finres}) leads to the following approximate formula  
for the energy dependence of charged multiplicity in central $Au-Au$ collisions:
$$
\left<{2 \over N_{part}}\ {d N_{ch} \over d \eta}\right>_{\eta < 1} \approx 0.87\ 
\left({\sqrt{s}\ ({\rm GeV}) \over 130}\right)^{0.25}\ \times \nonumber
$$
\be
\times \left[3.93 + 0.25\ \ln\left({\sqrt{s}\ ({\rm GeV}) \over 130}\right)
\right]. \label{endep}
\ee
At $\sqrt{s} = 130\ {\rm GeV}$, we estimate from Eq.(\ref{endep}) 
$2/N_{part}\ dN_{ch}/d\eta \mid_{\eta<1} = 3.42 \pm 0.15$, 
to be compared to the average experimental value of $3.37 \pm 0.12$ 
\cite{PHOBOS,Phenix,Star,Brahms}. 
At $\sqrt{s} = 200\ {\rm GeV}$, one gets $3.91 \pm 0.15$, 
to be compared to the PHOBOS value \cite{PHOBOS} of $3.78 \pm 0.25$. 
Finally, at $\sqrt{s} = 56\ {\rm GeV}$, we find $2.62 \pm 0.15$, 
to be compared to \cite{PHOBOS} $2.47 \pm 0.25$. 
It is interesting to note that formula (\ref{endep}), when extrapolated to very high energies, 
predicts for the LHC energy a value 
substantially smaller than found in other approaches:
\be
\left<{2 \over N_{part}}\ {d N_{ch} \over d \eta}\right>_{\eta < 1} = 10.8 \pm 0.5; \ \ \  \sqrt{s} = 5500\ {\rm GeV},
\ee
corresponding only to a factor of $2.8$ increase in multiplicity between the RHIC energy of  $\sqrt{s} = 200\ {\rm GeV}$ 
and the LHC energy of $\sqrt{s} = 5500\ {\rm GeV}$  
(numerical calculations show that when normalized to the number of participants, 
the multiplicity in central $Au-Au$ and $Pb-Pb$ systems is almost identical).
The energy dependence of charged hadron multiplicity per participant pair is shown in Fig.\ref{fig:mult}.

One can also try to extract the value of the exponent $\lambda$ from the energy dependence of hadron 
multiplicity measured by PHOBOS at $\sqrt{s} = 130\ \rm{GeV}$ and at $\sqrt{s} = 56\ \rm{GeV}$; 
this procedure yields $\lambda \simeq 0.37$, which is larger than 
the value inferred from the HERA data (and is very close to the value $\lambda \simeq 0.38$, 
resulting from the final--state saturation calculations \cite{EKRT}).

\subsubsection{Radiating the classical glue}

\begin{figure}[h]
\begin{minipage}{9.5cm}   
\begin{center}
\epsfysize=9.4cm
\leavevmode
\hbox{ \epsffile{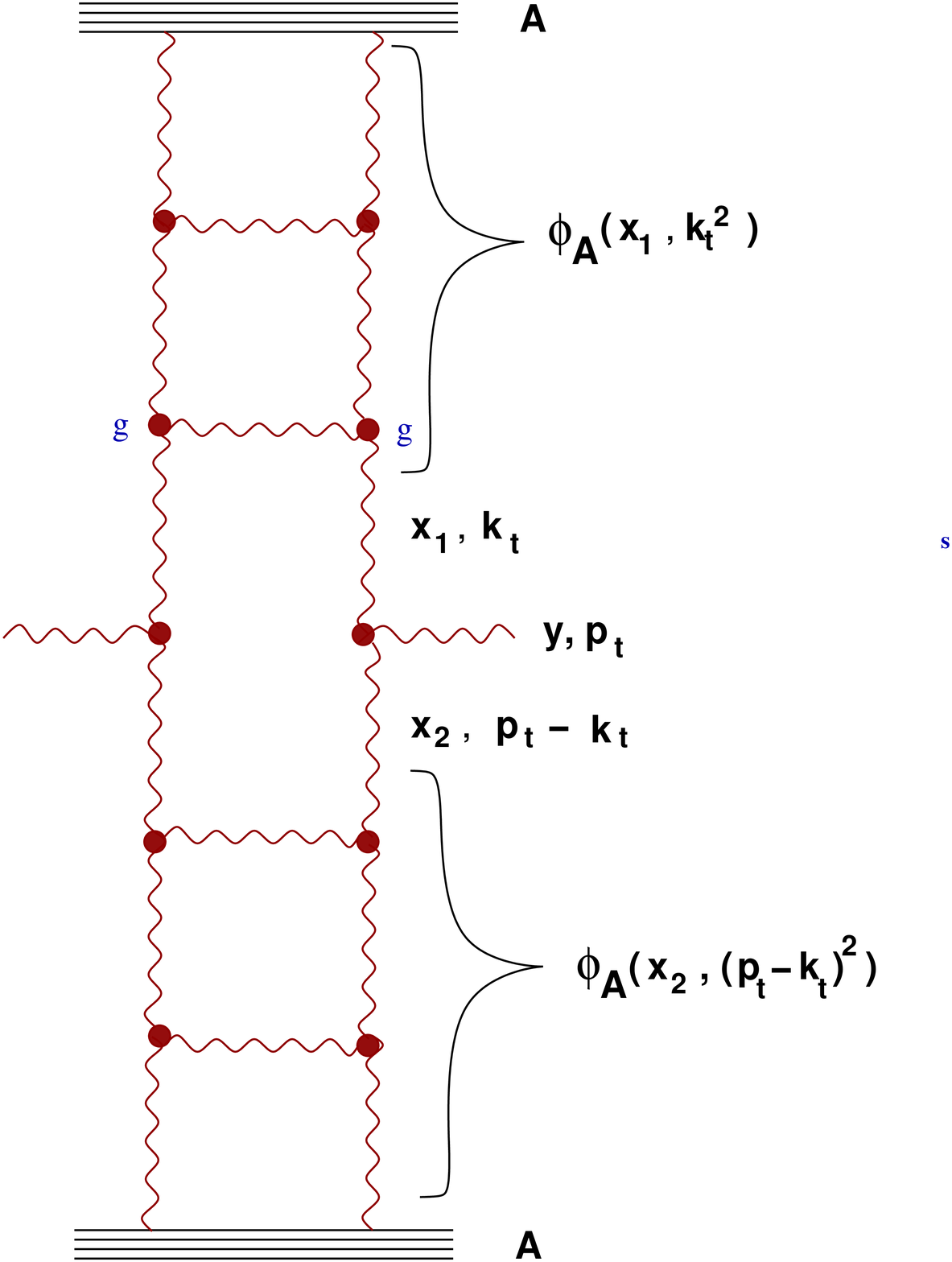}}
\end{center}
\end{minipage}
\begin{minipage}{5.5cm}
\caption{\em The Mueller diagram for the classical gluon radiation.}
\label{phi}
\end{minipage}
\end{figure}

Let us now proceed to the quantitative calculation of the (pseudo-) 
rapidity and centrality dependences 
\cite{KL1}. 
We need to evaluate the leading tree diagram describing 
emission of gluons on the classical level, see Fig. \ref{phi}\footnote{Note that this ``mono--jet'' 
production diagram makes obvious the absence of azimuthal correlations in the saturation regime 
discussed above, see Eq.~(\ref{abscor}).}.

Let us introduce the unintegrated gluon distribution $\varphi_A (x, k_t^2)$ which 
describes the probability to find a gluon with a given $x$ and transverse 
momentum $k_t$ inside the nucleus $A$. As follows from this definition, 
the unintegrated distribution is related to the gluon structure function by
\beq
xG_A(x, p_t^2) = \int^{p_t^2} d k_t^2 \ \varphi_A(x, k_t^2);
\eeq
when $p_t^2 > Q_s^2$, the unintegrated distribution corresponding to the bremsstrahlung 
radiation spectrum is 
\beq
\varphi_A(x, k_t^2) \sim {\alpha_s \over \pi} \ {1 \over k_t^2}.
\eeq 
In the saturation region, the gluon structure function is given by 
(\ref{glustr}); the corresponding unintegrated gluon distribution has only logarithmic dependence on the 
transverse momentum: 
\beq
\varphi_A(x, k_t^2) \sim {S_A \over \alpha_s}; \ k_t^2 \leq Q_s^2, \label{unint}
\eeq
where $S_A$ is the nuclear overlap area, determined by the atomic numbers of the 
colliding nuclei and by centrality of the collision.

 The differential cross section 
of gluon production in a $AA$ collision can now be written down as \cite{GLR,GM}
\beq
E {d \sigma \over d^3 p} = {4 \pi N_c \over N_c^2 - 1}\ {1 \over p_t^2}\ \int d k_t^2 \ 
\alpha_s \ \varphi_A(x_1, k_t^2)\ \varphi_A(x_2, (p-k)_t^2), \label{gencross}    
\eeq
where $x_{1,2} = (p_t/\sqrt{s}) \exp(\pm \eta)$, with $\eta$ the (pseudo)rapidity of the 
produced gluon; the running coupling $\alpha_s$ has to be evaluated at the 
scale $Q^2 = max\{k_t^2, (p-k)_t^2\}$. 
The rapidity density is then evaluated from (\ref{gencross}) according to 
\beq
{dN \over d y} = {1 \over \sigma_{AA}}\ \int d^2 p_t \left(E {d \sigma \over d^3 p}\right), 
\label{rapden}
\eeq
where $\sigma_{AA}$ is the inelastic cross section of nucleus--nucleus interaction.

Since the rapidity $y$ and Bjorken variable are related by $\ln 1/x = y$, 
the $x-$ dependence of the gluon structure function translates into the following 
dependence of the saturation scale $Q_s^2$ on rapidity:
\beq
Q_s^2(s; \pm y) = Q_s^2(s; y = 0)\ \exp(\pm \lambda y). \label{qsy}
\eeq

As it follows from (\ref{qsy}), the increase of rapidity at a fixed $W \equiv \sqrt{s}$ 
moves the wave function of one of the colliding 
nuclei deeper into the saturation region, while leading to a 
smaller gluon density in the other, which as a result can be 
pushed out of the saturation domain. Therefore, depending on the value of rapidity, 
the integration over the transverse momentum in Eqs. (\ref{gencross}),(\ref{rapden}) can be split in 
two regions: i) the region $\Lambda_{QCD} < k_t < Q_{s,min}$ in which the wave 
functions are both 
in the saturation domain; and ii) the region  $\Lambda << Q_{s,min} < k_t < Q_{s,max}$ in which 
the wave function of 
one of the nuclei is in the saturation region and the other one is not. 
Of course, there is 
also the region of $k_t > Q_{s,max}$, which is governed by the usual perturbative dynamics, 
but our assumption here is that the r{\^o}le of these genuine hard processes in the bulk 
of gluon production is relatively small; in the saturation scenario, 
these processes represent quantum fluctuations above the classical background. It is worth 
commenting that in the conventional mini--jet picture, this classical background is absent, 
and the multi--particle production is dominated by perturbative processes. 
This is the main physical difference between the two approaches; for the production 
of particles with $p_t >> Q_s$ they lead to identical results.    

To perform the calculation according to (\ref{rapden}),(\ref{gencross}) away from $y=0$ we need also 
to specify the behavior of the gluon structure function at large Bjorken $x$ (and out of 
the saturation region).  
At $x \to 1$, this behavior is governed by the QCD counting rules, $xG(x) \sim (1-x)^4$, so 
we adopt the following conventional form: $xG(x) \sim x^{-\lambda}\ (1-x)^4$.
  
We now have everything at hand to perform the integration over transverse momentum 
in (\ref{rapden}), (\ref{gencross}); the result is the 
following \cite{KL1}:
$$ 
{dN \over d y} = const\ S_A\ Q_{s,min}^2 \ \ln\left({Q_{s,min}^2 \over \Lambda_{QCD}^2}\right)
\ \times
$$
\beq
\times \ \left[1 + {1 \over 2}\ \ln\left({Q_{s,max}^2 \over Q_{s,min}^2}\right)\ 
\left(1 - {Q_{s,max} \over \sqrt{s}} e^{|y|}\right)^4\right],
 \label{resy}    
\eeq 
where the constant is energy--independent, $S_A$ is the nuclear overlap area, 
$Q_s^2 \equiv Q_s^2(s; y = 0)$, and $Q_{s,min(max)}$ 
are defined as the smaller (larger) values of (\ref{qsy}); at $y=0$, 
$Q_{s,min}^2 = Q_{s,max}^2 = Q_s^2(s) = Q_s^2(s_0)\ \times (s /s_0)^{\lambda / 2}$. 
The first term in the brackets in (\ref{resy}) originates from the region in which both 
nuclear wave functions are in the saturation regime; this corresponds to 
the familiar $\sim (1/\alpha_s)\ Q_s^2 R_A^2$ term in the gluon multiplicity. 
The second term comes from the region in which only one of the wave functions is in 
the saturation region. The coefficient $1/2$ in front of the second term in 
square brackets comes from $k_t$ ordering of gluon momenta in evaluation of 
the integral of Eq.(\ref{gencross}).

The formula (\ref{resy}) has been derived using the form (\ref{unint}) 
for the unintegrated gluon distributions. We have checked numerically that the use of 
more sophisticated functional form of  
$\varphi_A$ taken from the saturation model of Golec-Biernat and W{\"u}sthoff \cite{GW} 
in Eq.(\ref{gencross}) affects the results only at the level of about $3\%$.

\vskip0.3cm

Since $S_A Q_s^2 \sim N_{part}$ (recall that $Q_s^2 \gg  \Lambda_{QCD}^2$ is defined 
as the density of partons 
in the transverse plane, which is proportional to the density of participants), we can 
re--write (\ref{resy}) in the following final form \cite{KL1} 
$$
{dN \over d y} = c\ N_{part}\ \left({s \over s_0}\right)^{\lambda \over 2}\ e^{- \lambda |y|}\ 
\left[\ln\left({Q_s^2 \over \Lambda_{QCD}^2}\right) - \lambda |y|\right]\ \times
$$
\beq
\times \left[ 1 +  \lambda |y| \left( 1 - {Q_s \over \sqrt{s}}\ e^{(1 + \lambda/2) |y|} \right)^4 
\right],  
\label{finres}
\eeq
with $Q_s^2(s) = Q_s^2(s_0)\ (s /s_0)^{\lambda / 2}$.
This formula expresses the predictions of 
high density QCD for the energy, centrality, rapidity, and atomic number dependences 
of hadron multiplicities in nuclear collisions in terms of a single scaling function. 
Once the energy--independent constant $c \sim 1$ and $Q_s^2(s_0)$ are determined 
at some energy $s_0$, Eq. (\ref{finres}) contains no free parameters. 
At $y = 0$ the expression (\ref{resy}) coincides exactly with the one 
derived in \cite{KN}, and extends it to describe the rapidity and energy dependences.  

\subsubsection{Converting gluons into hadrons}

The distribution (\ref{finres}) refers to the radiated gluons, while what is measured in experiment 
is, of course, the distribution of final hadrons. We thus have to make an assumption about the 
transformation of gluons into hadrons. The gluon mini--jets are produced with a certain virtuality, 
which changes as the system evolves; the distribution in rapidity is thus not preserved. 
However, in the analysis of jet structure it has been found that the {\it angle} 
of the produced gluon is remembered by the resulting 
hadrons; this property of ``local parton--hadron duality'' (see \cite{Yuri} and references therein) 
is natural if one assumes 
that the hadronization is a soft process which cannot change the direction of the emitted radiation.  
Instead of the distribution in the angle $\theta$, it is more convenient to use the distribution in 
pseudo--rapidity $\eta = - \ln \tan (\theta /2)$.  
Therefore, before we can compare (\ref{resy}) to the data, we have to convert the rapidity distribution 
(\ref{finres}) into the gluon distribution in pseudo--rapidity. We will then assume that the gluon and hadron 
distributions are dual to each other in the pseudo--rapidity space. 

To take account of the 
difference between rapidity $y$ and the measured pseudo-rapidity $\eta$, we have to multiply (\ref{resy}) 
by the Jacobian of the $y \leftrightarrow \eta$ transformation;
a simple calculation yields
\beq    
h(\eta; p_t; m) = \frac{\cosh \eta}{\sqrt{\frac{  m^2  \,+\,p_t^2}{p_t^2}\,\,+\,\,\sinh^2 
\eta}}, \label{Jac}
\eeq
where $m$ is the typical mass of the produced particle, and $p_t$ is its typical transverse 
momentum.
Of course, to plot the distribution (\ref{finres}) as a function of pseudo-rapidity, one also 
has to express rapidity $y$ in terms of pseudo-rapidity $\eta$; 
this relation is given by 
\beq
y (\eta; p_t; m) = {1 \over 2}\  \ln\left[{{\sqrt{\frac{  m^2  \,+\,p_t^2}{p_t^2}\,\,+\,\,\sinh^2 \eta}} + \sinh \eta} \over 
{{\sqrt{\frac{  m^2  \,+\,p_t^2}{p_t^2}\,\,+\,\,\sinh^2 \eta}} - \sinh \eta}\right]; \label{yeta}
\eeq
obviously, $h(\eta; p_t; m) = {\partial y (\eta; p_t; m) / \partial \eta}$. 

We now have to make an assumption about the typical invariant mass $m$ of the gluon mini--jet. 
Let us estimate it by assuming that the slowest hadron in the mini-jet decay is 
the $\rho$-resonance, with energy $E_{\rho} = (m_{\rho}^2 + p_{\rho,t}^2 + p_{\rho,z}^2)^{1/2}$, 
where the $z$ axis is pointing along the mini-jet momentum.  
Let us also denote by $x_i$ the fractions of the gluon energy $q_0$ 
carried by other, fast, $i$ particles in the mini-jet decay. Since the sum of transverse (with respect 
to the mini-jet axis) momenta 
of mini-jet decay products is equal to zero, the 
mini-jet invariant mass $m$ is given by 
$$
m^2_{jet}\,\equiv\, m^2 = ( \sum_i x_i q_0 + E_{\rho})^2 - (\sum_i x_i
q_z + p_{\rho,z})^2 \,\simeq\,\,
$$
\begin{equation}
\label{decay}
 \,\simeq\,\,
2 \sum_i x_i  q_z \cdot ( m_{\rho,t} - p_{\rho,z}) \equiv 2 Q_s \cdot m_{eff},
\end{equation}
where $m_{\rho,t}=(m_{\rho}^2 + p_{\rho,t}^2)^{1/2}$.
In \eq{decay} we used that $\sum_i x_i =1$ and $q_0 \approx q_z = Q_s$.
Taking $p_{\rho,z} \approx p_{\rho,t}  \approx\,300$ MeV and $\rho$ mass, we obtain $m_{eff}
\approx 0.5\,$ GeV.

We thus use the mass $m^2 \simeq 2 Q_s m_{eff} \simeq Q_s \cdot 1$ GeV in Eqs.(\ref{Jac},\ref{yeta}).
 Since the typical transverse momentum of the produced 
gluon mini--jet is $Q_s$, we take $p_t = Q_s$ in (\ref{Jac}).
The effect of the 
transformation from rapidity to pseudo--rapidity is the decrease of multiplicity at 
small $\eta$ by 
about $25-30 \%$, leading to the appearance of the $\approx 10 \%$ dip in the pseudo--rapidity 
distribution in the 
vicinity of $\eta = 0$. 
 We have checked that the change in the value of the mini--jet mass by two times affects the 
Jacobian at central pseudo--rapidity to about $\simeq 10\%$, leading to $\sim 3\%$ effect on the 
final result.

\begin{figure}[htbp] 
\epsfig{file=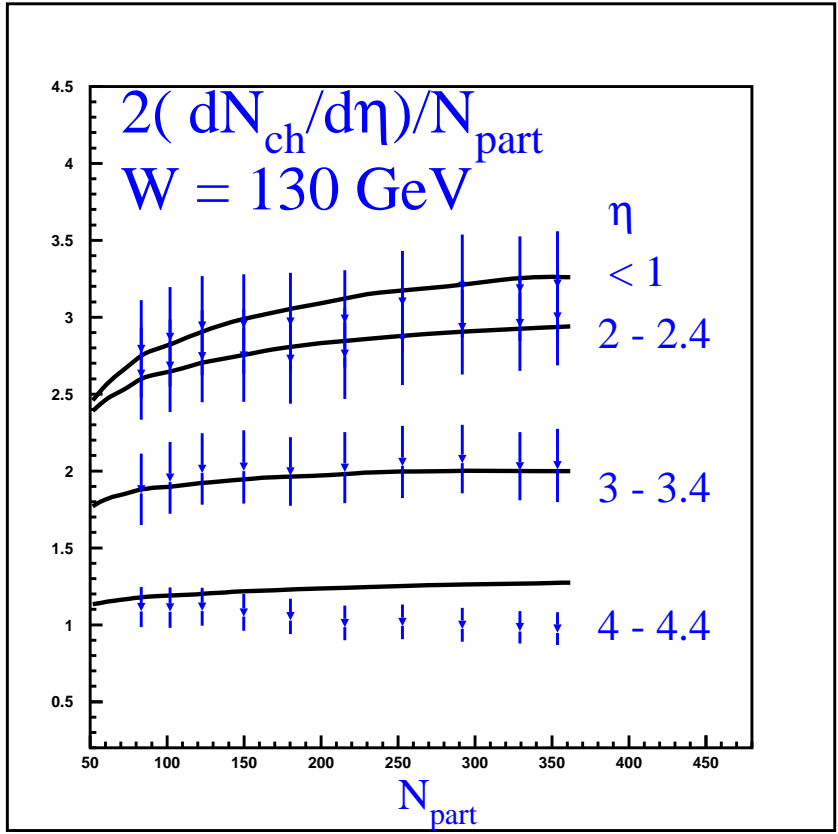,width=11cm}
\caption{\em
Centrality dependence of charged hadron production per participant at different 
pseudo-rapidity $\eta$ intervals in $Au-Au$ collisions 
at $\sqrt{s} = 130$ GeV; from \cite{KL1}, the data are from 
\cite{PHOBOS}.}
\label{fig1x}
\end{figure}

\begin{figure}[htbp] 
\epsfig{file=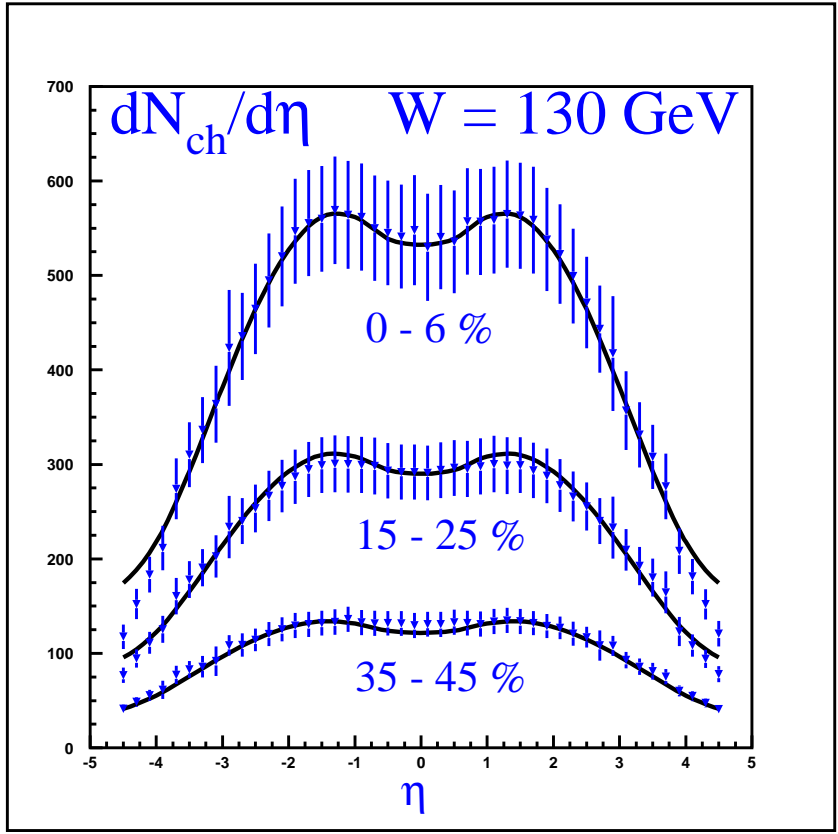,width=11cm}
\caption{\em
 Pseudo--rapidity dependence of charged hadron production at different cuts on centrality 
in $Au-Au$ collisions 
at $\sqrt{s} = 130$ GeV; from \cite{KL1}, the data are from 
\cite{PHOBOS}.}
\label{fig2}
\end{figure}

The results for the $Au-Au$ collisions at $\sqrt{s} = 130$ GeV are presented in Figs \ref{fig1x} 
and \ref{fig2}. In the calculation, we use the results on the dependence 
of saturation scale on the mean number of participants at $\sqrt{s} = 130$ GeV 
from \cite{KN}, see Table 2 of that 
paper. The mean number of participants in a given centrality cut is taken from the PHOBOS paper 
\cite{PHOBOS}. One can see that both the centrality dependence 
 and the rapidity dependence of the $\sqrt{s} = 130$ GeV PHOBOS data are well reproduced 
below $\eta \simeq \pm 4$. The rapidity dependence has been evaluated 
with $\lambda = 0.25$, which is within the 
range $\lambda = 0.25 \div 0.3$ inferred from the HERA data \cite{GW}. The discrepancy 
above $\eta \simeq \pm 4$ is not surprising since 
our approach does not properly take into account multi--parton correlations 
which are important in the fragmentation region.

Our predictions for $Au-Au$ collisions at $\sqrt{s} = 200$ GeV are presented in \cite{KL1}.
The only parameter which governs the energy dependence is the exponent $\lambda$, which we 
assume to be $\lambda \simeq 0.25$ as inferred from the HERA data. The absolute 
prediction for the multiplicity, as explained above, bears some uncertainty, but there is a definite 
feature of our scenario which is distinct from other approaches. 
It is the dependence of multiplicity on centrality, which around $\eta =0$ is determined 
solely by the running of the QCD strong coupling \cite{KN}. As a result, the centrality dependence 
at $\sqrt{s} = 200$ GeV is somewhat less steep than at $\sqrt{s} = 130$. While the 
difference in the shape at these two energies is quite small, in the perturbative 
mini-jet picture this slope 
should increase, reflecting the growth of the mini-jet cross section with 
energy \cite{WG}. 

\subsubsection{Further tests}

Checking the predictions of the semi--classical approach for the centrality and pseudo--rapidity 
dependence at $\sqrt{s} = 200$ GeV is clearly very important. What other tests 
of this picture can one devise? The main feature of the classical emission is that it is 
coherent up to the transverse momenta of about $\sqrt{2}\ Q_s$ (about $\simeq 2$ GeV/c for 
central $Au-Au$ collisions). This means 
that if we look at the centrality dependence of particle multiplicities above a certain value of the  
transverse momentum, say, above $1$ GeV/c, it should be very similar to the dependence 
without the transverse momentum cut-off. On the other hand, in the two--component ``soft plus hard'' model 
the cut on the transverse momentum would strongly enhance the contribution of hard 
mini--jet production processes, since soft production mechanisms presumably do not 
contribute to particle production at high transverse momenta.  
Of course, at sufficiently large value of the cutoff all of the observed particles 
will originate from genuine hard processes, and the centrality dependence will 
become steeper, reflecting the scaling with the number of collisions. It will be very 
interesting to explore the transition to this hard scattering regime experimentally. 

Another test, already discussed above (see Eq.~(\ref{abscor})) is the study of azimuthal 
correlations between the produced high $p_t$ particles. In the saturation scenario 
these correlations should be very small below $p_t \simeq 2$ GeV/c in central 
collisions. At higher transverse momenta, and/or for more peripheral collisions 
(where the saturation scale is smaller) these correlations should be much 
stronger.

\subsection{Does the vacuum melt?}

The approach described above allows us to estimate the initial energy density of partons 
achieved at RHIC. Indeed, in this approach the formation time of 
partons is $\tau_0 \simeq 1/Q_s$, and the transverse momenta of partons are about $k_t \simeq Q_s$. 
We thus can use the Bjorken formula and the set of parameters deduced above to estimate \cite{KN}
\be
\epsilon \simeq {<k_t> \over \tau_0} \ {d^2 N \over d^2b d \eta} \simeq Q_s^2 \ {d^2 N \over d^2b d \eta} \simeq 
18\ {\rm {GeV/fm^3}}
\ee
for central $Au-Au$ collisions at $\sqrt{s} = 130$ GeV.
This value is well above the energy density needed to induce the  
QCD phase transition according to the lattice calculations.  However, the picture of gluon production 
considered above seems to imply that the gluons simply flow from the initial state of the incident nuclei 
to the final state, where they fragment into hadrons, with 
nothing spectacular happening on the way. In fact, one may even wonder if the presence of these gluons 
modifies at all the structure of the physical QCD vacuum. 

To answer this question theoretically, we have to possess some knowledge about the non--perturbative 
vacuum properties. 
While in general the problem of vacuum structure still has not been solved (and this is one of the main 
reasons for the heavy ion research!), we do know one class of vacuum solutions -- the instantons.  
It is thus interesting to investigate what happens to the QCD vacuum in the presence of strong 
external classical fields using the example of instantons \cite{KKL}.

\begin{figure}
\epsfxsize=12cm
\leavevmode
\hbox{ \epsffile{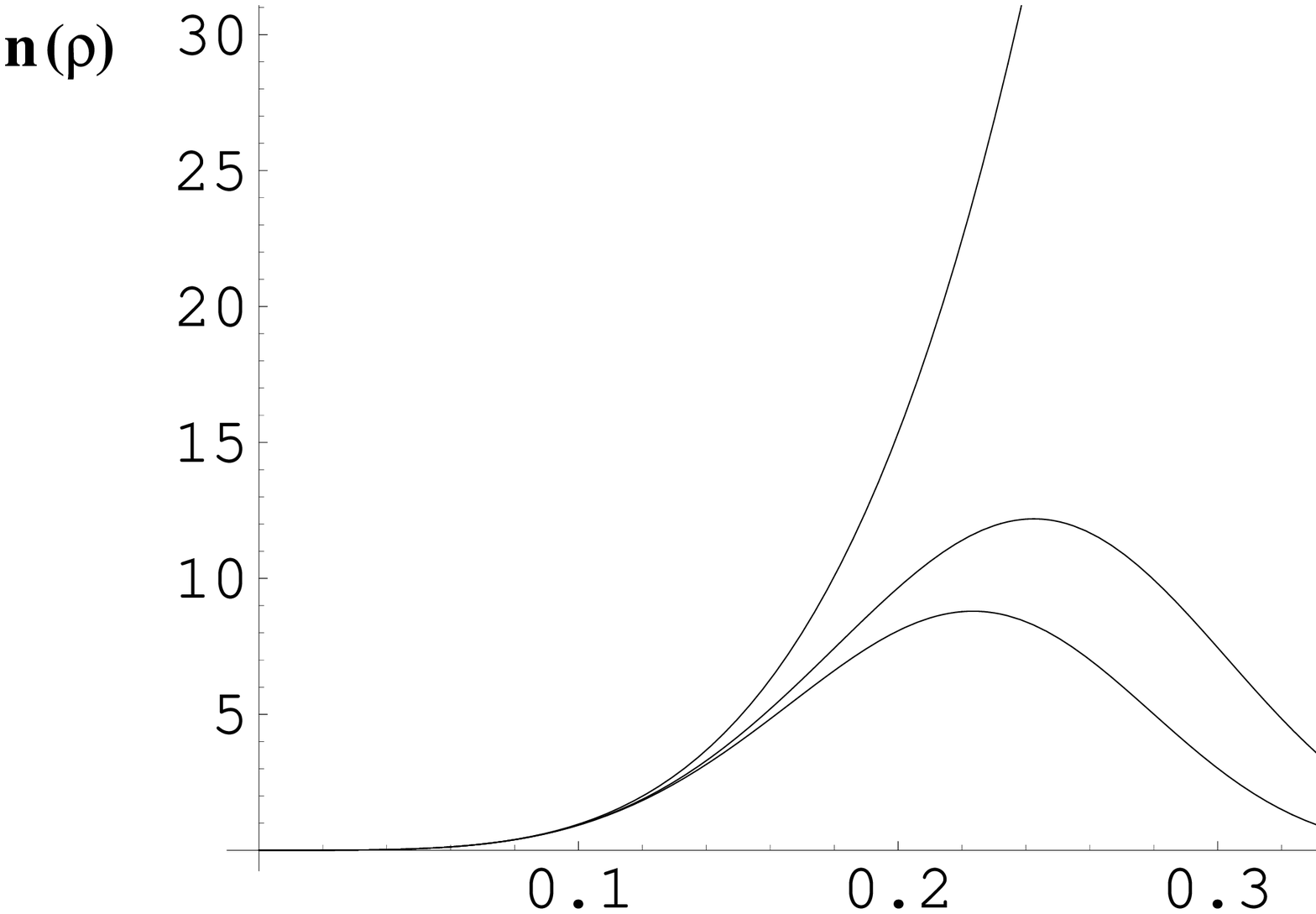}}
\caption{\em 
Distributions of instanton sizes in vacuum for QCD with 
three light flavors (upper curve) versus the distribution of instanton
sizes in the saturation environment produced by a collision of two
identical nuclei for $c=1$ (middle curve) and $c = 2 \ln 2$ (lower
curve) with $Q_s^2 \, = \, 2 \, \mbox{GeV}^2$; from 
\cite{inst}.}
\label{nsat}
\end{figure}

The problem of small instantons in a slowly varying background field
was first addressed in \cite{cdg1,svz} by introducing
the effective instanton Lagrangian $L^{I(\overline{I})}_{eff}
(x)$ 
\be\label{effl}
L_{eff}^I (x_0) \, = \, \int d \rho \, n_0 (\rho) \, dR \, \exp \left(
- \frac{2 \pi^2}{g^2} \, \rho^2 \, \overline{\eta}^M_{a\mu\nu} \, R^{a
a'} \, G^{a'}_{\mu\nu} (x_0) \right)
\ee
in which $n_0 (\rho)$ is the instanton size distribution function in the 
vacuum, $\overline{\eta}^M_{a\mu\nu}$ is the 't~Hooft symbol in Minkowski space, 
and $R^{aa'}$ is the matrix of rotations in color space, with $dR$ denoting 
the averaging over the instanton color orientations. 

 The complete field of a single instanton solution could be
reconstructed by perturbatively resumming the powers of the effective
instanton Lagrangian which corresponds to perturbation theory in
powers of the instanton size parameter $\rho^2$. In our case here the
background field arises due to the strong source current
$J_\mu^a$. The current can be due to a single nucleus, or resulting from the 
two colliding nuclei. Perturbative resummation of powers of
the source current term translates itself into resummation of the
powers of the classical field parameter $\as^2 A^{1/3}$
\cite{MV,yuri}. Thus the problem of instantons in the background 
classical gluon field is described by the effective action in
Minkowski space
\be\label{qcdi}
S_{eff} \, = \, \int d^4 x \left( - \frac{1}{4 g^2} G^a_{\mu\nu} (x)
G^a_{\mu\nu} (x) \, + \, L^I_{eff} (x) \, + \, L^{\overline{I}}_{eff}
(x) \, + \, J_\mu^a \, A_\mu^a (x) \right).
\ee

The problem thus is clearly formulated; by using an explicit form for the 
radiated classical gluon field, it was possible to demonstrate \cite{KKL} that
the distribution of
instantons gets modified from the original vacuum one $n_0(\rho)$ to
\be\label{naa}
n_{sat}^{AA} (\rho) \, = \, n_0 (\rho) \, \exp \left( - \frac{c \,
\rho^4 Q_s^4 } { 8 \, \as^2 \, N_c \, (Q_s \tau_0)^2 } \right),
\ee
where $\tau_0$ is the proper time.
The result \eq{naa} shows that large size instantons are 
suppressed by the strong classical fields generated in the nuclear
collision (see Fig. \ref{nsat})\footnote{Of course, at large proper times $\tau_0 \to \infty$ 
the vacuum ``cools off'', and the instanton distribution returns to the vacuum one.}.
The vacuum does melt!


\begin{theacknowledgments}
This manuscript is based on lectures given by one of us (D.K.) at the School on 
``New States of Matter in Hadronic Interactions'' of the Pan American Advanced Studies 
Institute in  Campos do Jord\~{a}o, S\~{a}o Paulo, Brazil,  on January 7-18, 2002. 
We thank the Organizers for the invitation to this stimulating meeting.

D.K. is grateful to his collaborators -- Yuri Kovchegov, Eugene Levin, 
and Marzia Nardi -- with whom most of the original results presented here were obtained.   
The work of D.K. was supported by 
the U.S. Department of Energy under Contract No. DE-AC02-98CH10886. He also 
wishes to acknowledge the hospitality of the Kavli Institute for Theoretical Physics 
in Santa Barbara, where this work was completed.  

J.R.\ wishes to thank the Gesellschaft f\"ur Schwerionenforschung (GSI),
Darmstadt, Germany, and the organizers 
for financial support during this summer school.
The work of J.R.\ was supported in part by the U.S.\ Department of Energy 
at Los Alamos National Laboratory under Contract No.\ W-7405-ENG-38.

We are grateful to Siggi Bethke and Hilmar Forkel for their kind permission
to reproduce two of their figures (Figs.~\ref{fig:bethke} and \ref{fig:vacuum})
in these lecture notes.

\end{theacknowledgments}





\end{document}